# A deterministic truthful PTAS for scheduling related machines

George Christodoulou[*]    Annamária Kovács[†]

**Abstract**

Scheduling on related machines ($Q||C_{\max}$) is one of the most important problems in the field of Algorithmic Mechanism Design. Each machine is controlled by a selfish agent and her valuation can be expressed via a single parameter, her *speed*. In contrast to other similar problems, Archer and Tardos [4] showed that an algorithm that minimizes the makespan can be truthfully implemented, although in exponential time. On the other hand, if we leave out the game-theoretic issues, the complexity of the problem has been completely settled — the problem is strongly NP-hard, while there exists a PTAS [9, 8].

This problem is the most well studied in single-parameter algorithmic mechanism design. It gives an excellent ground to explore the boundary between truthfulness and efficient computation. Since the work of Archer and Tardos, quite a lot of deterministic and randomized mechanisms have been suggested. Recently, a breakthrough result [7] showed that a randomized truthful PTAS exists. On the other hand, for the deterministic case, the best known approximation factor is 2.8 [11, 12].

It has been a major open question whether there exists a deterministic truthful PTAS, or whether truthfulness has an essential, negative impact on the computational complexity of the problem. In this paper we give a definitive answer to this important question by providing a truthful *deterministic* PTAS.

## 1 Introduction

*Algorithmic Mechanism Design (AMD)* is an area originated in the seminal paper by Nisan and Ronen [15, 16] and it has flourished during the last decade. It studies combinatorial optimization problems, where part of the input is controlled by selfish agents that are either unmotivated to report them correctly, or strongly motivated to report them erroneously, if a false report is profitable. In classical mechanism design more emphasis has been put on incentives issues, and less to computational aspects of the optimization problem at hand. On the other hand, traditional algorithm design disregards the fact that in some settings the agents might have incentive to lie. Therefore, we end up with algorithms that are fragile against selfish behavior. AMD carries challenges from both disciplines, aiming at the design of qualitative algorithms that, at the same time, give incentives to selfish users to report truthfully, and so are also immune to strategic behavior.

A fundamental optimization problem that has been suggested in [16] as a ground to explore the design of truthful mechanisms, is the *scheduling problem*, where a set of $n$ tasks need to be processed by a set of $m$ machines. There are two important variants with respect to the processing capabilities of the machines, that have been studied within the AMD framework. The machines can be *unrelated*, i.e., each machine $i$ needs $t_{ij}$ units of time to process task $j$; or *related*, where machine $i$ comes with a speed $s_i$, while task $j$ has processing requirement $p_j$, that is, $t_{ij} = p_j/s_i$ (we will use the settled notation $Q||C_{\max}$ to refer to the latter problem). The objective is to allocate the jobs to the machines so that the maximum finish time of the machines, i.e., the *makespan* is minimized.

---

[*]Max-Planck-Institut für Informatik, Saarbrücken, Germany. Email: gchristo@mpi-inf.mpg.de
[†]Department of Informatics, Goethe University, Frankfurt M., Germany. Email: panni@cs.uni-frankfurt.de

In the game-theoretic setting, it is assumed that each machine $i$ is a rational agent who controls the *private* values of row $t_i$. It is further assumed that each machine wants to minimize its completion time, and without any incentive it will lie, if this can trick the algorithm to assign less work to him. In order to motivate the machines to cooperate, we pay them to execute the tasks. A *mechanism* consists of two parts: an *allocation algorithm* that assigns the tasks to the machines, and a *payment scheme* that compensates the machines in monetary terms. We are interested in devising *truthful* mechanisms in *dominant* strategies, where each player maximizes his utility by telling the truth, regardless of the reports of the other players.

The scheduling problem provides an excellent framework to study the computational aspects of truthfulness. It is a well-studied problem from the algorithmic perspective with a lot of algorithmic techniques that have been developed. Moreover, it is conceptually close to combinatorial auctions, so that solutions and insights can be transferred from the one problem to the other. Indeed, the scheduling problem comes with a variety of objectives to be optimized, that are different than the objectives used in classical mechanism design.

From the traditional algorithmic point of view, the computational complexity of the related machines case problem is completely settled: There is a polynomial time approximation scheme (PTAS) [9] for an arbitrary number of machines, and an FPTAS [10] when the number of machines is fixed. The general case is strongly NP-complete, so we don't expect to find an FPTAS unless P=NP.

The mechanism design version of scheduling on related machines was first studied by Archer and Tardos [4]. It is the most central and well-studied among single-parameter problems, where each player controls a single real value and his objective is proportional to this value (see Chapters 9 and 12 of [14] for a precise definition). Myerson [13] gave a characterization of truthful algorithms for one-parameter problems, in terms of a monotonicity condition. Archer and Tardos [4] found a similar monotonicity characterization, and using it they showed that a certain type of optimal allocation is monotone and consequently truthful (albeit exponential-time).

The fact that truthfulness does not exclude optimality, in contrast to the multi-parameter variant of scheduling (the unrelated case)[1], makes the problem an appropriate example to explore the interplay between truthfulness and computational complexity. It has been a major open problem whether or not a *deterministic* monotone PTAS exists for $Q||C_{\max}$[2]. In this work, we give a definitive positive answer to that central question and conclude the problem.

## 1.1 Related Work

Auletta et al. [5] gave the first deterministic polynomial-time monotone algorithm for any fixed number of machines, with approximation ratio 4. This result was improved to an FPTAS by Andelman et al. [2]. For an arbitrary number of machines, Andelman, Azar, and Sorani [1] gave a 5-approximation deterministic truthful mechanism, and Kovács improved the approximation ratio to 3 [11] and to 2.8 [12], which was the previous record for the problem.

---

[1] With the scheduling on unrelated machines, we are more in the dark *(see [6] for a recent overview of results)*. There are impossibility results that show that there does not exist any truthful mechanism with approximation ratio better than a constant *even in exponential time*. Therefore, more primitive questions need to be answered before we settle the complexity of the problem. The only known algorithm for the problem is the VCG that has approximation ratio equal to the number of machines.

[2] We say that a mechanism runs in polynomial time when both the allocation algorithm and the payment algorithm run in polynomial time.

Randomization has been successfully applied. There are two major concepts of randomization of truthful mechanisms, *universal truthfulness*, and *truthfulness-in-expectation*. The first notion is strongest, and consists of randomized mechanisms that are probability distributions over deterministic truthful mechanisms. In the latter notion, by telling the truth a player maximizes his expected utility. Only the second notion of randomized truthfulness has been applied to the problem. Archer and Tardos [4] gave a truthful-in-expectation mechanism with approximation ratio 3, that was later improved to 2 [3]. Recently, Dhangwatnotai et al. [7], settled the status for the randomized version of the problem by giving a randomized PTAS that is truthful-in-expectation. Both mechanisms apply (among other methods) a randomized rounding procedure. Interestingly, randomization is useful only to guarantee truthfulness and has no implication on the approximation ratio. Indeed, both algorithms can be easily derandomized to provide deterministic mechanisms that preserve the approximation ratio, but violate the monotonicity condition.

## 1.2 Our results and techniques

We provide a deterministic monotone PTAS for $Q||C_{\max}$. The corresponding payment scheme [4] is polynomially computable[3], and with these payments our algorithm induces a $(1+3\epsilon)$-approximate deterministic truthful mechanism, settling the status of the problem.

We start by fixing a common basis for our subsequent considerations. We always assume that input speeds are indexed so that $s_1 \leq s_2 \leq \ldots \leq s_m$ holds. For any set of jobs $P = \{p_1, p_2, \ldots, p_j\}$, the *weight* or *workload* of the set is $|P| = \sum_{r=1}^{j} p_r$. We will view an allocation of the jobs to the machines as an *(ordered) partition* $(P_1, P_2, \ldots, P_m)$ of the jobs into $m$ *sets*. We search for an output where the workloads $|P_i|$ are in non-decreasing order.

The PTAS [8] – which is a simplified and polished version of the very first PTAS [9] – defines a directed network on $m + 1$ layers depending on the input job set, where each arc leading between the layers $i - 1$ and $i$ represents a possible realization of the set $P_i$, and directed paths leading over the $m$ layers correspond to the possible job partitions. An optimal solution is then found using a shortest path computation in this network.

The difficulty in applying any known PTAS to construct a deterministic monotone algorithm for $Q||C_{\max}$ is twofold. First, in all of the known PTAS's, sets of input jobs of approximately the same size form groups, s.t. in the optimization process a common (rounded or smoothed) size is assumed for all members of the same group. Second, jobs that are tiny compared to the total workload of a machine do not turn up individually in the calculations, but just as part of an arbitrarily divisible (e.g., in form of small blocks) total volume.

Note that it must be relatively easy to find an allocation procedure that is in a way 'approximately monotone'. However, (exact) monotonicity intuitively requires exact determination and knowledge of the allocated workloads. To illustrate this, we just point out that in every monotone (in expectation) algorithm for $Q||C_{\max}$ provided so far, the (expected) workloads either occur in increasing order wrt. increasing machine speeds, or constitute a lexicographically minimal optimal solution wrt. a fixed solution set and a fixed machine indexing.

Thus, both of the mentioned simplifications of the input set – which, to some extent, seem necessary to admit polynomial time optimization – appear to be condemned to destroy any attempt to make a deterministic adaptation monotone. (The authors of [7] used randomization at both points to obtain the monotone in expectation PTAS.) Our ideas to eliminate the above two sources of inaccuracy of the output are the following, respectively:

1. As for rounding the job sizes, note that grouping is necessary only to reduce the (exponential)

---
[3]This is intuitively clear, since our *work curve* is a step function with a polynomial number of steps.

number of different outputs. We can achieve the same goal if for any group of jobs of similar size we fix the order of jobs in which they appear in the allocation (e.g., in increasing order), *and*, calculate with the exact job sizes along the optimization process. Notice that not even the fact is obvious that such a solution with increasing workloads exists. Now, if reducing a machine speed increases the makespan of the (previously optimal) solution, that means that this machine became a bottleneck, so a new upper bound on the optimum makespan over the considered set of outputs is induced *exactly* by the (previous) workload of the changed machine (the same argument as used in [4, 2, 7]). With this idea we derandomize the first type of randomization (*job smoothing*) of [7].

2. Concerning tiny jobs, we observe that with these we can fill up some of the fastest machines nearly to the makespan level. On the other hand, it is easy to show [3] that pre-rounding machine speeds to powers of some predefined $(1+\epsilon)$ does not spoil monotonicity and increases the approximation bound by only a factor of $(1+\epsilon)$. Assuming now that the coarsity of tiny blocks is *much* finer than the coarsity of machine speeds, we can be sure that (full) machines of higher speed receive more work than slower machines. Moreover, having reduced the speed of such a machine, tiny jobs in its workload 'flow' to other machines to provide a makespan 'much' smaller than implied by the previous workload of this machine.

It is quite a technical challenge to combine these two ideas so smoothly that in the end yields a correct monotonicity proof. We accomplish this task as follows. We fix (for the proof argument) a set $L_i$ of non-tiny jobs on each machine, so that the $L_1, L_2, \ldots, L_m$ have increasing and exactly known weights, and they fulfil the constraints suggested in 1. On top of the sets $L_i$, each machine has a set $S_i$ of small jobs (due to necessary conditions for rounding the total volume of tiny jobs, some of these are uniform blocks, while some are known exactly). The total set of small jobs is flexible (along the proof), in particular we can always move a small job to a higher index machine, and obtain a valid schedule. Moreover, we set the objectives so that in an optimum solution the small jobs are moved to the higher index machines as much as possible (and so, make them full).

Our monotonicity proof becomes subtle in case of the first (and so, not necessarily full) machine containing small jobs. It is especially so when this first machine is $m$, not leaving space for manipulating the small jobs in the output as needed. In order to circumvent this problem we restrict the search to allocations where at least two machines do have some tiny blocks (unless too few tiny jobs exist). Moreover, it seems crucial in our monotonicity argument that every machine has the possibility to get rid of all the tiny blocks (i.e., those inducing uncertain workload) if this is provoked by a reduction of its speed. Combining these two requirements we treat the last *three* machines as a single entity. A carefully optimized assignment of an 'obligatory' set of tiny blocks, and later of the actual tiny jobs to these machines then implies monotonicity.

## 1.3 Preliminaries

The input is given by a set $P_I$ of $n$ input jobs, and a vector $s$ (or $\sigma$) of input speeds $s_1 \leq \ldots \leq s_m$. For a job $p \in P_I$ we use $p$ both to denote the individual job, and the *size* of this job in a given formula. For a desired approximation bound $1 + \epsilon$, we choose a $\delta \ll \epsilon$, that will be the rounding precision of the job sizes. For ease of exposition, we will assume that $(1+\delta)^t = 2$ for some $t \in \mathbb{N}$.[4] Furthermore, we define $\rho$ as the unique integer power of 2 in $[\delta/6, \delta/3]$. We use the interval notation for a set of non-negative integers like, e.g., $[1, m]$. This should cause no confusion, as in such cases it will always be obvious that we consider integers.

---

[4]This assumption is unrealistic for computations, but it is *not* necessary for the result to hold. We could equally well use the rounding function of [8] or [7]. However, this would overload the paper with clumsy technicalities, e.g., in Definition 2. Also, since our result is of purely theoretical interest, we do not try to optimize the ratio $\delta/\epsilon$; it will be clear that, e.g., $30\delta < \epsilon$ suffices in the proofs.

**Definition 1** (job classes). *If $p$ denotes (the size of) a job, then $\overline{p}$ denotes this job rounded up to the nearest integral power of $(1+\delta)$. A job $p$ is in the job class $C_l$, iff $\overline{p} = (1+\delta)^l$.*

*Let $C_l = \{p_{l1}, p_{l2}, \ldots, p_{ln_l^{\max}}\}$ be the jobs of $C_l$ in some fixed non-decreasing order of size. We use the notation $C_l(a) = \{p_{l1}, \ldots, p_{la}\}$ for $0 \leq a \leq n_l^{\max}$, and $C_l(a, b) = C_l(b) \backslash C_l(a)$ for $0 \leq a \leq b \leq n_l^{\max}$.*

*If $P = \{p_1, p_2, \ldots, p_j\}$ is a job set, then $\overline{P} = \{\overline{p}_1, \overline{p}_2, \ldots, \overline{p}_j\}$ denotes the corresponding set of rounded jobs. The* weight *or* workload *of $P$ is $|P| = \sum_{r=1}^{j} p_r$; the* rounded weight *is $|\overline{P}| = \sum_{r=1}^{j} \overline{p}_r$. Assuming that the jobs are in* non-increasing *order of size, we denote the subset of the $r$ largest jobs by $P^r = \{p_1, p_2, p_3, \ldots, p_r\}$.*

## 2 Canonical allocations

This section characterizes the type of allocations – we call them *canonical allocations* – that we will consider. Definitions 2 and 3 describe the necessary restrictions on the (output) job partition $P_1, \ldots, P_m$. Subsequently, as our first main result, Theorem 1 states that for any input, and any $\delta > 0$, a canonical allocation exists that provides a $1+3\delta$ approximation to the optimum makespan.

**Definition 2** ($\delta$-division). *We say that a given set of jobs $P$ is $\delta$-divided into the pair of sets $(L, S)$ (or $P = (L, S)$) if*

(D1) $P = L \cup S$ *and* $L \cap S = \emptyset$,

(D2) $\overline{p} > \frac{\delta \cdot |L|}{(1+\delta)^2}$ *for every* $p \in L$, *and*

(D3) $\overline{q} \leq \delta |L|$ *for every* $q \in S$.

**Definition 3** (canonical allocation). *For a given input, an allocation $P_1, P_2, \ldots P_m$ is called* canonical*, if for every $i \in [1, m]$, the set $P_i$ can be $\delta$-divided into $(L(P_i), S(P_i))$ (or $(L_i, S_i)$, for short), so that the following properties hold:*

(A1) *If $i < i'$, then $|L_i| \leq |L_{i'}|$.*

(A2) *for jobs $p$ and $q$ of the same job class $p \leq q$ holds if and only if*

   (a) $p \in L_i$ *and* $q \in S_{i'}$ *for some* $i, i' \in [1, m]$, *or*
   (b) $p \in L_i$ *and* $q \in L_{i'}$ *and* $i \leq i'$, *or*
   (c) $p \in S_i$ *and* $q \in S_{i'}$ *and* $i \leq i'$.

In the proof of Theorem 1, we modify an optimal partition of the rounded input jobs $\overline{P}_I$ to get the canonical allocation: First we take the *core* set of each set in the partition (see Definition 4), then we order the sets by increasing order of core size, and apply Lemma 1 to make the modified cores fulfil property (A2) (b). It is easy to show that *small* jobs (those outside the cores) can be shifted to fast machines, where they remain small, and so still induce a $\delta$-division on each machine. First, we start with the definition of the *core*, and then we proceed with Lemmata 1 and 2, that are important ingredients of the proof of Theorem 1.

**Definition 4.** *Given a set $P$ of jobs, we define the* core *$cr(P)$ of $P$ as follows. Consider the jobs $P = \{p_1, p_2, \ldots\}$ in a fixed non-increasing order of size. Let $j$ be minimum with the property that $\overline{p}_j \leq \frac{\delta}{1+\delta} |P^{j-1}|$, then $cr(P) \stackrel{\text{def}}{=} P^{j-1} = \{p_1, \ldots, p_{j-1}\}$. If no such $j$ exists, then $cr(P) \stackrel{\text{def}}{=} P$.*

**Lemma 1.** *Let $(Q_1, \ldots, Q_m)$ be a partition of a subset $Q$ of the input jobs such that $|\overline{Q}_1| \leq \ldots \leq |\overline{Q}_m|$. There exists a partition $(L_1, \ldots, L_m)$ of $Q$ that satisfies:*

1. $|L_1| \leq \ldots \leq |L_m|$

2. *for any job class $C_l$, if job $p_{lj}$ belongs to $L_i$ and job $p_{lk}$ to $L_{i'}$ where $i < i'$, then $j < k$.*

3. *for all $i$*
$$\frac{1}{1+\delta}|\overline{Q}_i| < |L_i| \leq |\overline{Q}_i|.$$

*Proof.* Let $Q_i \subset R$. We say that we *maximize $Q_i$ wrt. $R$*, if for every class $l$ we replace the jobs in $Q_i \cap C_l$ by the largest possible jobs in $R \cap C_l$ (i.e., if there are $r$ such jobs then with the $r$ largest jobs of $R \cap C_l$). We will denote the maximized set by $Q_i^R$. Clearly, if $S \subset R$, then $|Q_i^S| \leq |Q_i^R|$.

Now we construct the new partition recursively. We define $L_m$ as a set of maximum workload among $\{Q_1^Q, Q_2^Q, \ldots, Q_m^Q\}$ (notice that the latter is not a partition of a subset of the input jobs). Assuming that $L_m = Q_i^Q$, now $L_{m-1}$ is defined to be a set of maximum workload among $\{Q_1^{Q \setminus L_m}, \ldots, Q_{i-1}^{Q \setminus L_m}, Q_{i+1}^{Q \setminus L_m} \ldots, Q_m^{Q \setminus L_m}\}$, etc. In every recursive step we selected a set that has larger weight than any other remaining set, even if those sets get the largest remaining jobs of the respective classes. This proves 1., whereas 2. holds by construction.

Next we argue that 3. holds as well. Observe that $\{\overline{Q}_1, \overline{Q}_2, \ldots, \overline{Q}_m\}$ and $\{\overline{L}_1, \overline{L}_2, \ldots, \overline{L}_m\}$ (as sets) are exactly the same. The proof of
$$\frac{1}{1+\delta}|\overline{Q}_i| < |L_i| \leq |\overline{Q}_i|$$
is simply the fact that there exist at least $i$ jobsets among the $\overline{Q}_l$, so that $|\overline{Q}_l| < (1+\delta)|L_i|$, (namely, the sets of rounded jobs $\overline{L}_1, \overline{L}_2, \ldots, \overline{L}_i$), on the other hand there exist at least $m - i + 1$ sets among the $\overline{Q}_l$, so that $|L_i| \leq |\overline{Q}_l|$ (namely, the sets $\overline{L}_i, \overline{L}_{i+1}, \ldots, \overline{L}_m$). □

**Lemma 2.** *Let $P$ be a set of jobs, then*

*(c1)* $\forall p \in cr(P) \quad \overline{p} > \frac{\delta}{(1+\delta)^2}|cr(P)|$;

*(c2)* $\forall q \in P \setminus cr(P) \quad \overline{q} \leq \frac{\delta}{1+\delta}|cr(P)|$.

*Proof.* (c2) is trivially true, since job sizes are non-increasing. By Definition 4 it holds that $\overline{p}_{j-1} > \frac{\delta}{(1+\delta)}|P^{j-2}|$. Therefore

$$(1+\delta)^2 \overline{p}_{j-1} > (1+\delta)\overline{p}_{j-1} + \delta p_{j-1} > \delta(|P^{j-2}| + p_{j-1}) = \delta|P^{j-1}| = \delta|cr(P)|.$$

The same holds for all jobs not smaller than $p_{j-1}$, which proves (c1). □

**Theorem 1.** *For arbitrary increasing input speeds and input jobs, a canonical allocation inducing a schedule with makespan at most $(1 + 3\delta)OPT$ exists, where $OPT$ is the optimum makespan of the input.*

*Proof.* Let $P$ be the set of all jobs and $s_1 \leq \ldots \leq s_m$ be the input speeds. We process this set of jobs in five steps to finally obtain the desired canonical allocation. In the next two steps we consider only the set of rounded jobs $\overline{P}$.

*1. (core division)* We start from an optimal schedule of $\overline{P}$. Let this be $(P_1, P_2, \ldots, P_m)$ and its makespan be $M \leq (1+\delta)OPT$. The inequality trivially holds, since any schedule of $P$ induces a schedule of $\overline{P}$ of makespan increased by a factor of at most $(1+\delta)$.

Moreover, for every $P_i$ let $S_i = P_i \setminus cr(P_i)$. In the rest of the proof we call jobs in $\bigcup_{i=1}^m cr(P_i)$ *large*, and jobs in $\bigcup_{i=1}^m S_i$ *small*.

*2. (core sorting)* In this step we start from the schedule $(P_1, P_2, \ldots, P_m)$ with $P_i = cr(P_i)\dot\cup S_i$ and we result in a (fractional) schedule $P'_1, \ldots, P'_m$ with $P'_i = L'_i\dot\cup S'_i$, where $L'_1, L'_2, \ldots, L'_m$ is simply the set of cores $cr(P_i)$ sorted by weight. Each small job might be cut into finitely many parts, and distributed over the sets $S'_i$. Importantly, $P'$ has makespan at most $M$.

We define the rearranged sets $S'_i$ of small jobs in course of sorting $\{P_i\}$ step by step, with insertion: after step $i$, $cr(P_1), cr(P_2), \ldots, cr(P_i)$ become sorted by weight, and the jobs of $\bigcup_{h=1}^i S_h$ are allocated fractionally to machines $1, 2, \ldots i$, so that the makespan remains $M$, and the sets $P_{i+1}, \ldots, P_m$ remain intact.

Now we explain how we redistribute the small jobs. When we insert the set $P_i = cr(P_i)\dot\cup S_i$ to some position $k \leq i$, then the job sets previously on machines $k, k+1, \ldots, i-1$ move to the next higher index machine, where they clearly fit below $M$. Even though all the jobs in $P_i$ might not fit on machine $k$ (below $M$), certainly the jobs of $cr(P_i)$ do. This is because the workload that was previously on machine $k$, had a coreset larger than $cr(P_i)$. Moreover, notice that *all* jobs previously (before step $i$) on machines $k, k+1, \ldots, i$ altogether fit on the same set of machines below $M$.

Now we don't move large jobs at all, but take the small jobs in the same order as they are allocated now, starting from (small) jobs on machine $i, \ldots, k$, and continuously 'fill' them to the machines in the same decreasing order of the machines, cutting a (fractional) job into two when the time $M$ is reached. (Alternatively, we can just pick the superfluous jobs of $S_i$, and fill them (fractionally) to empty gaps of machines $k+1, \ldots, i$. )

*Observation.* Every (fractional) small job that was previously in $S_i$, now moves to a $P'_h$, with $|L'_h| \geq |cr(P_i)|$. This implies that (c2) of Lemma 2 still holds (a fractional job fulfils (c2), if its original full size does). Furthermore, (c1) trivially holds, since we did not change the large sets.

*3. (permutation)* Now we return to the original jobsizes. We replace the rounded jobs in each $L'_i$ by original jobs, so that we use the smallest possible jobs within in every class. We want that (A1) and (A2) hold, so we apply Lemma 1 on the resulting partition. After applying the procedure of Lemma 1, we obtain $L_1, \ldots, L_m$. By the lemma we know that $\frac{1}{1+\delta}|L'_i| < |L_i| \leq |L'_i|$ holds for every $i$. This implies, on the one hand, that the makespan is still at most $M$. On the other hand, (leaving the $S'_i$ as they were) we obtained $\delta$-divisions $(L_i, S'_i)$ : (D2) holds, since for any job $p$ in $L_i$ we have by (c1) that

$$\overline{p} > \frac{\delta}{(1+\delta)^2}|\overline{L}_i| \geq \frac{\delta}{(1+\delta)^2}|L_i|;$$

(D3) holds, because if $q \in S'_i$ then by (c2)

$$\overline{q} < \frac{\delta}{(1+\delta)}|L'_i| \leq \delta|L_i|.$$

*4. (small job sorting)* Finally, we fill the small jobs continuously on the machines below $M$, in decreasing order of size (still fractional allocation), starting from machine $m$. This ensures (A2)(c), and we claim that it does not spoil (D3) either: if, due to this sorting, now some job $q$ were too large for the set $L_i$ of the machine, then that would mean that *all* the small jobs of size at least $q$ must have been on machines $i+1, \ldots, m$, (while the large sets were the same), and fit below $M$, which is impossible as shown by the ordered allocation.

5. *(integral allocation)* We make an integral allocation by assigning every fractional job to the fastest machine where the job occurs. Note that by the previous construction, every machine gets at most one such job. This increases the makespan to at most $(1+\delta)M \leq (1+\delta)^2 OPT$, because we had (and still have) $\delta$-divisions. □

## 3 Configurations

Like in [8, 7], we introduce so called *configurations* $\alpha(w, \mu, \vec{n}^o, \vec{n}^1)$ in order to represent any possible job set $P_i$ of the partition, up to $\delta$ accuracy. We use the configurations to define the vertices of a directed graph $\mathcal{H}$. A well-defined optimal path in this graph will then specify our output schedule.[5]

The first component of any configuration is a *magnitude $w$* which is an integer power of 2. As we proceed from slow machines to fast machines in a schedule, the monotonically increasing magnitude keeps track of the largest job size allocated so far, which must be some size in the interval $(w/2, w]$. Thus, the current magnitude also shows, which (larger) job sizes are not yet relevant, and which (tiny) jobs need not be taken into account *individually* anymore in the configuration. This motivates the next definition.

**Definition 5** (valid magnitude)**.** *The value $w = 2^z$ ($z \in \mathbb{Z}$) is a* valid magnitude *if an input job $p \in P_I$ exists so that $w/2 < p \leq w$. Let $w_{\min}$ and $w_{\max}$ denote the smallest and the largest valid magnitudes, respectively. We call a job* tiny *for $w$ if it has size at most $\rho w$.*

Recall that $\rho$ is the integer power of 2 between $\delta/6$, and $\delta/3$. Having a magnitude $w$ fixed, let $\lambda = \log_{(1+\delta)} \rho w = t \cdot \log(\rho w)$, and $\Lambda = \log_{(1+\delta)} w = t \cdot \log w$, where $(1+\delta)^t = 2$. Notice that both $\lambda$ and $\Lambda$ are integers, and by Definition 1, the jobs of size in $(\rho w, w]$ belong to the classes $C_{\lambda+1}, \ldots, C_\Lambda$. These will constitute the relevant job classes, if the largest jobsize on the current or slower machines is between $w/2$ and $w$.

If the configuration $\alpha$ represents the set $P_i$ in a job partition, then the so-called *size vector* $\vec{n}^o = (n^o_\lambda, n^o_{\lambda+1}, \ldots, n^o_\Lambda)$ describes the jobs in the cumulative job set $A_{i-1} \stackrel{\text{def}}{=} \bigcup_{h=1}^{i-1} P_h$ as follows. For $\lambda < l \leq \Lambda$, $(l \neq \mu, \mu+1)$, exactly the first (smallest) $n^o_l$ jobs of the class $C_l$ are in the set $A_{i-1}$. Moreover, in $A_{i-1}$ the total weight of jobs from $\bigcup_{l \leq \lambda} C_l$ is in the interval $((n^o_\lambda - 1) \cdot \rho w, (n^o_\lambda + 1) \cdot \rho w)$. However, the particular subset of these small jobs inside $A_{i-1}$, is not determined by $\alpha$. The vector $\vec{n}^1$ represents the set $A_i \stackrel{\text{def}}{=} \bigcup_{h=1}^{i} P_h$, in the same way.

A major difference to the configurations of [8], is that our configurations should not only represent a job set $P_i$, but also its $\delta$-division $(L_i, S_i)$. In particular, we will distinguish four types of job sizes in a configuration. *Tiny* jobs have size at most $\rho w$, and, as already seen, are represented by the first coordinates $n_\lambda$ of the two size vectors with their total size rounded to an integer multiple of $\rho w$. Correspondingly, we will sometimes talk about *blocks* of size $\rho w$ which are simply re-tailored tiny jobs for the purposes of our analysis.

**Definition 6.** Blocks *are imaginary tiny jobs, each having size $\rho w$ for some valid magnitude $w$. We use $S(n_\lambda, \rho w)$ to denote a set of $n_\lambda$ blocks of size $\rho w$.*

*Small* jobs are those that (together with the tiny jobs), can only appear in the set $S_i$ of the $\delta$-division, whereas *large* jobs can only be in the set $L_i$. However, there must exist job classes – we will call them *middle* size jobs –, which might occur in both $L_i$ and $S_i$, since by (D2) and (D3) of Definition 2 there is a flexible border between the job sizes in $L_i$ and in $S_i$. Therefore, exactly two job classes, $\mu$ and $\mu + 1$ will be represented by a *triple* of (increasing) non-negative integers, $\underline{n}_\mu = (n_{\mu\ell}, n_{\mu m}, n_{\mu s})$, and $\underline{n}_{(\mu+1)} = (n_{(\mu+1)\ell}, n_{(\mu+1)m}, n_{(\mu+1)s})$, instead of scalar values $n_\mu$, and $n_{(\mu+1)}$ in both of the vectors $\vec{n}^o, \vec{n}^1$. In the case of $\underline{n}^o_\mu$, the meaning of the three numbers will be that in the set $A_{i-1}$, from the job class $C_\mu$ exactly the jobs in $C_\mu(n^o_{\mu m}, n^o_{\mu s})$ are allocated as small jobs, that is, to one of the sets $S_1, S_2, \ldots, S_{i-1}$, and exactly the jobs in $C_\mu(n^o_{\mu\ell})$ as large jobs, i.e., in one of $L_1, \ldots, L_{i-1}$, and similarly in case of $\vec{n}^1_\mu$ for the set $A_i$. The sets $C_\mu(n^o_{\mu s}, n^o_{\mu n^{\max}_\mu})$ and $C_\mu(n^o_{\mu\ell}, n^o_{\mu m})$ are to be allocated as small and large jobs, respectively, on higher index machines. The meaning of the numbers for $\mu + 1$ are analogous. Now we finished the preparations for the next two definitions.

---
[5]Roughly speaking, our graph can be thought of as the *line graph* of the graph $G$ defined in [8] (with simple modifications). That is, the vertices of $\mathcal{H}$ correspond to edges of $G$. This is the reason why our configurations include two vectors $\vec{n}_1$ and $\vec{n}_2$ instead of only one.

**Definition 7** (size vector). *A size vector $\vec{n} = (n_\lambda, \ldots, n_\Lambda)$ with middle size $\mu \in [\lambda+1, \Lambda]$, is a vector of integers, with the exception of the entries $\underline{n}_\mu = (n_{\mu\ell}, n_{\mu m}, n_{\mu s})$ and $\underline{n}_{\mu+1} = (n_{(\mu+1)\ell}, n_{(\mu+1)m}, n_{(\mu+1)s})$ both of which are themselves vectors of three integers, so that $n_{\mu\ell} \leq n_{\mu m} \leq n_{\mu s}$, and $n_{(\mu+1)\ell} \leq n_{(\mu+1)m} \leq n_{(\mu+1)s}$ holds. All of the integer entries belong to $[0, n]$.*

**Definition 8** (configuration). *A configuration $\alpha(w, \mu, \vec{n}^o, \vec{n}^1)$ consists of four components: a valid magnitude $w$, and two size vectors $\vec{n}^o = (n^o_\lambda, \ldots, n^o_\Lambda)$, and $\vec{n}^1 = (n^1_\lambda, \ldots, n^1_\Lambda)$ with middle size $\mu$, such that*

(C1) $n^o_l \leq n^1_l \leq n^{\max}_l$ *for $\lambda < l \leq \Lambda$, $l \notin \{\mu, \mu + 1\}$;*

(C2) *if $w \neq w_{\min}$ then $n^1_l > 0$ for at least one $l \in (\Lambda - t, \Lambda]$;*

(C3) $n^o_\lambda \leq n^1_\lambda \leq \left\lceil \frac{\sum_{l \leq \lambda} |C_l|}{\rho w} \right\rceil + 3;$

(C4) $n^o_{\mu\ell} \leq n^1_{\mu\ell} \leq n^o_{\mu m} = n^1_{\mu m} \leq n^o_{\mu s} \leq n^1_{\mu s} \leq n^{\max}_\mu$, *and analogously for $\mu + 1$;*

$$T_\alpha \stackrel{\text{def}}{=} S(n^1_\lambda - n^o_\lambda, \rho w).$$

$$L_\alpha \stackrel{\text{def}}{=} C_\mu(n^o_{\mu\ell}, n^1_{\mu\ell}) \cup C_{(\mu+1)}(n^o_{(\mu+1)\ell}, n^1_{(\mu+1)\ell}) \cup \bigcup_{l=\mu+2}^{\Lambda} C_l(n^o_l, n^1_l), \text{ and}$$

$$S_\alpha \stackrel{\text{def}}{=} C_\mu(n^o_{\mu s}, n^1_{\mu s}) \cup C_{(\mu+1)}(n^o_{(\mu+1)s}, n^1_{(\mu+1)s}) \cup \bigcup_{l=\lambda+1}^{\mu-1} C_l(n^o_l, n^1_l) \cup T_\alpha.$$

(C5) *either $\vec{n}^o \neq \vec{n}^1$, and $(1+\delta)^{(\mu+1)} \leq \delta \cdot |L_\alpha| < (1+\delta)^{(\mu+2)}$;*

*or $\alpha$ is the* empty configuration *$(w_{\min}, \lambda_{\min} + 1, \vec{0}, \vec{0})$ where $\lambda_{\min} = t \cdot \log(\rho w_{\min})$.*

**Notation.** *We refer to the whole represented job set $L_\alpha \cup S_\alpha$ (including virtual blocks) simply by $\alpha$ (abusing notation), and $|\alpha|$ stands for the total work of the set $\alpha$. We denote the set without tiny blocks by $\tilde{\alpha} = \alpha \setminus T_\alpha$.*

It is easy to verify, that the requirements (C1), (C3) and (C4) are necessary, if we want $\vec{n}^o$ and $\vec{n}^1$ to represent cumulative job-sets of a partition the way we described above. (C2) implies that $w$ is always the smallest possible magnitude for representing these job-sets. (C5) is different in flavor from the previous four properties: it implicates that the set $L_\alpha$ and $\mu$ strongly affect each-other.

However, it can be shown (as we do in proving Theorem 2) that for every set $P_i = A_i \setminus A_{i-1}$ (and corresponding $w$) in a canonical schedule a unique $\mu > \lambda$ exists that fulfils (C5).

We stress here that the cumulative sets $A_i$ do *not* possess a $\delta$-division, and a single size vector $\vec{n}$ does not represent an $(L, S)$ division at all. On the other hand, any configuration, indeed, represents a $\delta$-division (see Lemma 3).

## 4 The directed graph $\mathcal{H}_I$

In this section, for arbitrary input instance $I$, we define a directed, layered graph $\mathcal{H}_I$. All vertices of this graph are configurations, selected, numbered, and 'chained' to form the graph in an appropriate way.

First, in Section 4.1, for an arbitrary configuration $\alpha$, we define a set $Scale(\alpha)$ of configurations. These are the possible configurations of an end-vertex of any arc with a starting vertex having $\alpha$ as configuration. We took the name *scale* from [8], where $scale_{w \to w'}(\vec{n}) = \vec{n}'$ is a single size vector that represents the same set of jobs as $\vec{n}$ does, from the aspect of some higher magnitude $w'$ than $w$. Similarly, in our case, if $\alpha = (w, \mu, \vec{n}^o, \vec{n}^1)$, $\beta = (w', \mu', \vec{n}'^o, \vec{n}'^1)$, and $\beta \in Scale(\alpha)$, then $\vec{n}'^o$ must represent the same job set $A_i$, as $\vec{n}^1$, from the point of view of a (possibly) increased magnitude $w'$ and a (possibly) increased middle size $\mu'$. Next, in Section 4.2, we proceed with the exact defition of the graph, and finally in Section 4.3, we prove that a minimum shortest path in this graph, corresponds to an allocation with approximation ratio $1 + \mathcal{O}(\delta)$.

### 4.1 The definition of $Scale(\alpha)$

The exact definition of $Scale()$ might look somewhat technical. Nevertheless, this is mainly due to the middle sizes $\mu$ and $\mu'$. Disregarding (S2), the conditions below are the natural 'scaling requirements', as also appeared in [8]. In the definition we will use the notation $\lambda = \log_{(1+\delta)}(\rho w)$, $\Lambda = \log_{(1+\delta)} w$, $\lambda' = \log_{(1+\delta)}(\rho w')$, and $\Lambda' = \log_{(1+\delta)} w'$.

**Definition 9** ($Scale(\alpha)$). *Let $\alpha = (w, \mu, \vec{n}^o, \vec{n}^1)$, and $\beta = (w', \mu', \vec{n}'^o, \vec{n}'^1)$ be two configurations, where $\vec{n}^1 = \vec{n} = (n_\lambda, \ldots, n_\Lambda)$, resp. $\vec{n}'^o = \vec{n}' = (n'_{\lambda'}, \ldots, n'_{\Lambda'})$; then $\beta \in Scale(\alpha)$ iff*

(S1) $w \leq w'$, and $\mu \leq \mu'$;

(S2) *if $\mu' = \mu$ then $\underline{n}'_\mu = \underline{n}_\mu$ and $\underline{n}'_{\mu+1} = \underline{n}_{\mu+1}$; if $\mu < \mu'$, then $n_{\mu\ell} = n_{\mu m}$ and $n'_{(\mu'+1)m} = n'_{(\mu'+1)s}$ ;*

*if $\mu + 1 = \mu'$ then $\underline{n}_{(\mu+1)} = \underline{n}'_{\mu'}$; if $\mu + 1 < \mu'$, then $n_{(\mu+1)\ell} = n_{(\mu+1)m}$ and $n'_{\mu'm} = n'_{\mu's}$ ;*

*For the sake of a concise presentation, in the next three requirements we assume that $n_\mu \stackrel{\text{def}}{=} n_{\mu s}$, and $n'_{(\mu'+1)} \stackrel{\text{def}}{=} n'_{(\mu'+1)\ell}$, whenever $\mu < \mu'$ holds, furthermore $n_{(\mu+1)} \stackrel{\text{def}}{=} n_{(\mu+1)s}$, and $n'_{\mu'} \stackrel{\text{def}}{=} n'_{\mu'\ell}$, if additionally $\mu + 1 < \mu'$ holds.*

(S3) *if $\Lambda < l \leq \Lambda'$, then $n'_l = 0$;*

(S4) *if $\lambda' < l \leq \Lambda$, then $n_l = n'_l$*

(S5) *If $n_\lambda = 0$, then let $n'_{\lambda'} = \left\lceil \frac{\sum_{l=\lambda+1}^{\lambda'} |C_l(n_l)|}{\rho w} \right\rceil$. Otherwise let $\tau_\alpha = n_\lambda \rho w + \sum_{l=\lambda+1}^{\lambda'} |C_l(n_l)|$, and $n'_{\lambda'}$ be the smallest nonnegative integer such that*

$$(\tau_\alpha - \rho w, \tau_\alpha + \rho w) \subset ((n'_{\lambda'} - 1) \cdot \rho w', (n'_{\lambda'} + 1) \cdot \rho w').$$

We provide some intuition concerning Definition 9, by comparing $\vec{n}$ and $\vec{n}'$, the old and the new size vectors, respectively. First of all note that if $\alpha$ and $\beta$ represent the consecutive sets $P_i$ and $P_{i+1}$ of a partition, then, indeed, both of these vectors should represent the same cumulative job set $A_i = \sum_{h=1}^{i} P_h$.

Besides the natural demand of increasing $w$ and $\mu$, which we keep in its simplest form (S1), the 'traditional' scaling requirements are (S3) to (S5). By (S3) and (S4), job classes not appearing in $\vec{n}$ do not occur in $A_i$, whereas those appearing explicitly in both $\vec{n}$ and $\vec{n}'$ must be represented by the same number in both size vectors.

Less obvious is (S5). By the first condition we want to achieve that $n'_{\lambda'} = 0$ if and only if *no* jobs of size at most $\rho w'$ have been allocated in $A_i$. Now – by inductive argument – the total size of tiny jobs in $A_i$ must be between $(n_\lambda - 1)\rho w$ and $(n_\lambda + 1)\rho w$. During scaling to $w'$ we shift this interval by the exact workload of jobs that become tiny right now, and so obtain the interval $(\tau_\alpha - \rho w, \tau_\alpha + \rho w)$. Now $n'_{\lambda'} \cdot \rho w'$ has to be the midpoint of a new, (longer) interval containing $(\tau_\alpha - \rho w, \tau_\alpha + \rho w)$ as a subset. For $w' = w$ we clearly obtain $\tau_\alpha = n_\lambda \rho w$, and so $n_\lambda = n'_{\lambda'}$. Assume now that $w\rho = 1$, and $w'\rho = 2$. Observe that any interval of length 2 (i.e., $(\tau_\alpha - \rho w, \tau_\alpha + \rho w)$), either contains an integer multiple of 2 or has it as a (lower) endpoint. This will be $n'_{\lambda'} \cdot 2 = n'_{\lambda'} \cdot \rho w'$, the middle of the larger interval (here of length 4) that covers the original interval completely. Since the new interval can also be covered by a properly positioned interval of length 8, and so on, this proves that also for $\rho w' = 4, 8, 16\ldots$, etc., an appropriate $n'_{\lambda'}$ exists. Here we exploited that the magnitudes, and $\rho$ are exact powers of 2.

Finally, we turn to the meaning of (S2). As long as $\mu$ remains a middle size in the new size vector $\vec{n}'$, the same triple $\underline{n}_\mu$ represents the set of jobs allocated in $A_i$ as small resp. as large jobs, from the class $C_\mu$. If $\mu$ becomes smaller than the new middle size $\mu'$, that means that the jobs of $C_\mu(n_{\mu m})$, that have to be allocated as large jobs, *have* already been allocated, that is, $n_{\mu \ell} = n_{\mu m}$ and so $C_\mu(n_{\mu m}) = C_\mu(n_{\mu \ell}) \subseteq A_i$. Moreover, $C_\mu(n_{\mu s})$ are now *all* the jobs allocated from this class, so we can define (the scalar) $n_\mu \stackrel{\text{def}}{=} n_{\mu s}$. Similarly, if $\mu'$ is not a middle size in the old vector $\vec{n}$, then no jobs of class $C_{\mu'}$ have been allocated as small up to the set $A_i$, and this is expressed by $n'_{\mu' m} = n'_{\mu' s}$ , and by the notation $n'_{\mu'} \stackrel{\text{def}}{=} n'_{\mu' \ell}$. The considerations for $\mu + 1$ and for $\mu' + 1$ are analogous.

## 4.2 The graph $\mathcal{H}_I$

The vertices of $\mathcal{H}_I$ (i.e., the configurations) are arranged in $m$ *layers*, and in *levels I* and *II*, which are orthogonal to the layers. The configurations on level $I$ must have an empty set of small jobs, i.e., $S_\alpha = \emptyset$, and here the layers $\{m-2, m-1, m\}$ are empty. Level II has $m-1$ 'real' layers, and we add a single dummy vertex $v_m$ adjacent to every vertex on layer $m-1$, that alone forms the last layer $m$.[6] In general, the $i$th layer stands for the $i$th set $P_i$. Any directed path of $m$ nodes leads to $v_m$ over the $m$ layers, and from level $I$ (or $II$) to level $II$. Such a path we will call an *m-path*. The $m$-paths will represent partitions of the input $P_I$.

For a given $m$-path, the very first vertex on level II is in some layer $k \leq m-2$; we will call it the *switch vertex*, and $k$ the *switch machine* or *switch index*. (Note that $k$ is thus the first machine possibly receiving small jobs.) We shall denote the vertices on the two levels by $V_I$, and $V_{II}$, respectively.

**Notation.** *For any directed path $(v_1, v_2, \ldots v_r)$, the corresponding configurations of the nodes will be denoted by $(\alpha_1, \alpha_2, \ldots, \alpha_r)$.*

---

[6]More precisely, we will unite layers $m-2$ and $m-1$, and use *double vertices* in the united layer, but it simplifies the discussion to think of these as pairs of individual vertices.

At this point, let us briefly discuss about the last set $P_m$ of the partition. Note that for an $m$-path, the last configuration $\alpha_{m-1}$ alone represents $\bigcup_{h=1}^{m-1} P_h$. Thus, we can use $\alpha_{m-1}$ to uniquely determine the 'hidden configuration' $\alpha_m$ (not appearing explicitely in the path). We define $\alpha_m$ as follows: let $w_m = w_{m-1}$, $\mu_m = \mu_{m-1}$, and $\vec{n}^o$ be the second size vector in $\alpha_{m-1}$ extended by 0 entries on $(\Lambda_{m-1}, \Lambda_{\max}]$. Furthermore, $n_\lambda^1 = \left\lceil \frac{\sum_{l \leqslant \lambda} |C_l|}{\rho w} \right\rceil + 3$; $n_{ls}^1 = n_l^{\max}$ and $n_{l\ell}^1 = n_{lm}^1$ for $l \in \{\mu, \mu+1\}$, and $n_l^1 = n_l^{\max}$ if $l \notin \{\lambda, \mu, \mu+1\}$. Observe, that $\alpha_m$ represents *all* jobs of class higher than $\Lambda_{m-1}$ (the algorithm can handle this job set as a huge chunk without violating the running time bounds).[7] Keeping $w_{m-1} = w_m$, plays an important role in the monotonicity proof.

Furthermore, also due to the monotonicity requirement, we want to handle even the last *three* workloads $\alpha_{m-2}, \alpha_{m-1}$, and $\alpha_m$ together. In particular, we will require that either all of them have the same magnitude, and therefore use $w'_{m-1} = w_{m-2}$ instead of $w_{m-1}$, or that $w_{m-2}$ is *much* smaller than $w_{m-1}$, so that all jobs on $m-2$ (if exist), are tiny for machines $m-1$ and $m$ (cf. cases (A) and (B) below).

We define the graph so that every $m$-path should represent a *canonical* allocation, as defined in Section 2. Beyond that, we impose further restrictions on the paths that we consider; these restrictions can also be reflected in the graph definition. Moreover, for a given speed vector, any $m$-path will have a naturally defined *makespan* value. Among the $m$-paths adhering to the restrictions, the algorithm selects an $m$-path having *minimum* makespan, as the primary objective. Among paths of minimum makespan, we maximize the index of the switch machine $k$. A further order of preference, and restriction to be of type (A) or (B) is the following. Observe that in case (A) on the last three machines, resp. in case (B) on the last two machines the block-size for tiny jobs is the same.

(A) $w_{m-2} > \rho^2 \cdot w_{m-1}$; in this case we modify the last magnitude to be $w'_{m-1} := w_{m-2}$, and require $|\tilde{\alpha}_{m-2}| \leq |\tilde{\alpha}_{m-1}| \leq |\tilde{\alpha}_m|$, and $|\alpha_{m-2}| \leq |\alpha_{m-1}| \leq |\alpha_m|$; moreover,

  (i) either *all* tiny jobs (measured by blocks) are on machines $m-1$ and $m$, or

  (ii) machines $\{m-2, m-1, m\}$ have at least 18 tiny blocks, and at least two of them have each at least 6 tiny blocks.

(B) $w_{m-2} \leq \rho^2 \cdot w_{m-1}$; then $|\tilde{\alpha}_{m-1}| \leq |\tilde{\alpha}_m|$, and $|\alpha_{m-1}| \leq |\alpha_m|$, and

  (i) all machines but $\{m-1, m\}$ are empty, or

  (ii) $m-1$ and $m$ together have at least 6 tiny blocks.

The requirements (A) and (B) can be incorporated in the graph, e.g., by using (polynomially many) special *double vertices* $v'_{m-2}$ with *double configurations* $(\alpha_{m-2}, \alpha_{m-1})$ on level $II$. Applying $w'_{m-1} := w_{m-2} > \rho^2 \cdot w_{m-1}$ can be done by using size vectors of triple length for the double vertices of type (A). Clearly, all restrictions can be represented by the configurations $(\alpha_{m-2}, \alpha_{m-1})$.

The subsequent definition of graph $\mathcal{H}_I$ is independent of the speed vector $s$, and depends only on the job set $P_I$. After that, we assign a weight to each vertex, called *finish time*, and define the *makespan* of a path accordingly. Obviously, these values do depend on the machine speeds $s$.

We assume, w.l.o.g. that $m \geq 3$, otherwise we include a machine of speed 0.

**Definition 10** (graph $\mathcal{H}_I$). *$\mathcal{H}_I(V, E)$ is a directed graph, where every vertex $v \neq v_m$ is a triple $v = (d, i, \alpha)$, so that $d \in \{I, II\}$, $i$ is an integer in $[1, m-1]$, and $\alpha = (w, \mu, \vec{n}^o, \vec{n}^1)$ is a configuration. In particular, each triple that obeys the rules (V1) to (V4) below, determines a vertex in $V$.*

---

[7]Because of $\mu_{m-1} = \mu_m$, we can only require $(1+\delta)^{(\mu+1)} \leq \delta \cdot |L_{\alpha_i}|$ instead of property (C5) of configurations. As a consequence, on the last machine the division $(L_m, S_m)$ does not fulfil (D2).

(V1) *if $i = 1$, then for $l \neq \mu, \mu+1$ $n^o_l = 0$, while for $l \in \{\mu, \mu+1\}$ $n^o_{l\ell} = 0$ and $n^o_{lm} = n^o_{ls}$;*

(V2) *if $d = I$, then $i \in [1, m-3]$, and $S_\alpha = \emptyset$;*

(V3) *if $d = II$, then $i \in [1, m-3]$, and $n^1_\lambda \leq \max\{n^o_\lambda, \left\lfloor \frac{\sum_{l \leq \lambda} |C_l|}{\rho w} \right\rfloor - 1\}$;*

*There is an arc from $v = (d, i, \alpha)$ to $v' = (d', i+1, \beta)$ if an only if*

(E1) $\beta \in Scale(\alpha)$, *and*

(E2) $|L_\alpha| \leq |L_\beta|$, *and*

(E3) $d \leq d'$;

(V4) *Finally, for $d = II$, and combined layers $(m-2, m-1)$, include double vertices $v'$ with double configurations $(\alpha_{m-2}, \alpha_{m-1})$ so that for $(\alpha_{m-2}, \alpha_{m-1}, \alpha_m)$ the requirements (E1) (E2) and either (A) or (B) hold.*

**Definition 11** (finish time of a vertex). *Let $v = (d, i, \alpha)$ be a vertex of $\mathcal{H}_I$, where $\alpha = (w, \mu, \vec{n}^o, \vec{n}^1)$. The* finish time *of $v$ is then $f(v) = \frac{|\alpha| + \rho w}{s_i}$ if $n^o_\lambda < n^1_\lambda$, and $f(v) = \frac{|\alpha|}{s_i}$ otherwise.*

**Definition 12** (makespan of a path). *Let $\mathcal{Q} = (v_1, v_2, \ldots, v_r)$ be a directed path in $\mathcal{H}_I$. If $\mathcal{Q} \subset V_I$, or $\mathcal{Q} \subset V_{II}$, then the* makespan *of $\mathcal{Q}$ is $M(\mathcal{Q}) = \max_{h=1}^r f(v_h)$. If $\mathcal{Q}$ is an $m$-path with switch vertex $v_k = (II, k, \alpha_k)$, then $M(\mathcal{Q}) = \max\{\frac{|\alpha_k|}{s_k}, \max_{h \neq k} f(v_h)\}$.*

Definition 12 allows $v_r = v_m$. The finish time $f(v_m)$ is calculated from the hidden configuration $\alpha_m$, as uniquely determined by $v_{m-1}$.

## 4.3 Approximation ratio of minimum-cost path

Theorem 2, saying that an $m$-path having (path-)makespan close to the optimum makespan of the scheduling problem always exists, is a consequence of Theorem 1. The proof is rather straightforward, and requires a technical translation of real schedules to $m$-paths of $\mathcal{H}_I$, which involves creating blocks of size $\rho w_i$ from the actual tiny jobs. In order to prove Theorem 2, we will make use of the following two technical lemmata.

**Lemma 3.** *For any configuration $\alpha$, the sets $(L_\alpha, S_\alpha)$ form a $\delta$-division of the set $\alpha$.*

*Proof.* The smallest job that might occur in $L_\alpha$ is at least from the class $C_\mu$, therefore $\overline{p} \geq (1+\delta)^\mu$ for any $p \in L_\alpha$. This implies $\overline{p} \cdot (1+\delta)^2 \geq (1+\delta)^{\mu+2} > \delta \cdot |L_\alpha|$, by the property (C5), and thus, we obtained (D2) for $(L_\alpha, S_\alpha)$.

Similarly, for any $q \in S_\alpha$, we have $\overline{q} \leq (1+\delta)^{\mu+1}$. This proves (D3), since $(1+\delta)^{\mu+1} \leq \delta \cdot |L_\alpha|$, by (C5). Obviously, $L_\alpha \cap S_\alpha = \emptyset$, and $\alpha = L_\alpha \cup S_\alpha$, so (D1) holds, and $(L_\alpha, S_\alpha)$ is, indeed, a $\delta$-division. □

Lemma 3 implies that tiny blocks in any $\alpha_i$ are small wrt. $|L_{\alpha_i}|$. The following observation sets a more exact bound on the block size.

*Observation.* For any (non-empty) configuration $\alpha_i = (w, \mu, \vec{n}^o, \vec{n}^1)$ in an $m$-path,

$$\frac{3}{2}\rho w < \delta |L_{\alpha_i}|. \tag{1}$$

The observation holds, since $w/2 < |L_{\alpha_h}|$ for some $h \leq i$, according to (C2), (E1) and the requirements of $Scale()$. By (E2), $|L_{\alpha_h}| \leq |L_{\alpha_i}|$. Now we have $\rho < \delta/3$, and $w/2 < |L_{\alpha_i}|$.

In the following proof(s), we will frequently say we 'put' small jobs from one machine to another, although we are actually modifying some $m$-path $\mathcal{Q}$ to obtain another path $\mathcal{Q}'$. Technically, this can be done as follows. Suppose that we put one job from class $l \geq \lambda$ from machine $i$ to $i+1$, where $\alpha(w, \mu, \vec{n}^o, \vec{n})$, and $\beta(w', \mu', \vec{n}', \vec{n}^1)$ with smallest job-classes $\lambda$ and $\lambda'$ are the configurations of $v_i$ and $v_{i+1}$, respectively. In $\alpha$ we reduce $n_l$ by 1. As for $\beta$, if $l > \lambda'$, then we reduce $n'_l$ by 1, and we are done. If $l \leq \lambda'$, then we scale the reduced $\vec{n}$ size vector according to (S5) in order to obtain the new $n'_{\lambda'}$. In this way, $n'_{\lambda'}$ either reduces by 1, or keeps its original value. In the latter case we can say that $T_\beta$ *swallowed* the job. If we put a job from $i$ onto $h > i$, we can repeatedly apply the above changes to the configurations, until the job gets swallowed, or we arrive at machine $h$. After such an act (given that we obtain a canonical allocation again), we arrive at an allocation that is represented by a corresponding $m$-path in the graph.

**Theorem 2.** *For every input $I = (P_I, s)$ of the scheduling problem, the optimal makespan over all $m$-paths in $\mathcal{H}_I$ is at most $OPT \cdot (1 + \mathcal{O}(\delta))$, where $OPT$ denotes the optimum makespan of the scheduling problem.*

*Proof.* Recall that $w_{\max}$ is the largest valid magnitude. We collect tiny jobs from $P_I$ into a set $T$ starting from the smallest job, and proceeding in increasing order of job size. We stop collecting, if either $|T| \geq 18\rho w_{\max}$, or the next job has size more than $\rho w_{\max}$. Let $P^o := P_I \setminus T$.

According to Theorem 1, a canonical allocation $P^o_1, \ldots, P^o_m$, $P^o_i = (L_i, S_i)$ of the jobs in $P^o$ exists with makespan of at most $OPT(1 + 3\delta)$. We modify this allocation $P^o_1, \ldots, P^o_m$ step by step, and finally obtain an appropriate path in $\mathcal{H}_I$.

First, we shift small jobs in $S_{m-2} \cup S_{m-1} \cup S_m$ to the right so that $|\tilde{\alpha}_{m-2}| \leq |\tilde{\alpha}_{m-1}| \leq |\tilde{\alpha}_m|$ holds, increasing the makespan by at most $3\delta OPT$.

Now we define the magnitude $w_i$ for each $i < m$ to be the smallest power of 2 that is at least $\max\{p \mid p \in \bigcup_{h=1}^i P^o_i\}$. For $m$ let $w_m := w_{m-1}$. Because of (A1), now $|L_i| > w_i/2$, and inequality (1) holds for $L_i$. In turn, (D2) and (1) imply that jobs of size at most $\rho w_i$ can only be in $S_i$ (and not in $L_i$). (Note that (D2) and (D3) admit that the largest job in $\bigcup_{h=1}^i P^o_i$ appears in some $S_i$. However, the previous sentence implies that it cannot belong to $T_i$. Therefore, after the subsequent modification, it remains in its original set $P^o_i$; that is, the defined magnitudes $w_i$ remain consistent.)

As the next step, we allocate the set $T$ of tiny jobs to the fastest machine, increasing the workload of $m$ by at most $18\rho w_{\max}$, and the makespan by at most $12\delta \cdot OPT$, (cf. (1)). Moreover, if $w_{m-2} \leq \rho^2 w_{m-1}$, then either all machines $i \leq m-2$ are empty, or $m-2$ is non-empty, meaning that $T$ has at least $18\rho w_{\max} \geq 18\rho w_{m-1}$ jobs of size at most $w_{m-2}$, which are jobs tiny for $w_{m-1}$, so (B) holds. If $w_{m-2} > \rho^2 w_{m-1}$, then either $T$ contains all jobs tiny for $w_{m-2}$, so that (A)(i) holds, or $T$ has enough tiny jobs so that (A)(ii) holds. In the latter case we distribute $T$ over machines $m-1$ and $m$. Let us denote the current *partition* of $P_I$ by $P_1, P_2, \ldots, P_m$.

In what follows, we modify this partition so that it contains an integer number of tiny blocks instead of the tiny jobs for every $i \neq m$.

Let $T_i = \{p \in P_i | p \leq \rho w_i\}$ be the set of tiny jobs in $P_i$. The jobs in $T_i$ make $|T_i|/(\rho w_i)$ (fractional) blocks. We can build integral blocks out of these for every $i < m$ by a simple procedure – also described in [7] – which packs (possibly) fractional tiny jobs from a fractional block on some machine $i$ into a fractional block on machine $h > i$, until one of them gets rounded to an integer number of blocks. Note that the (full size of) any repacked job remains tiny on its new machine. We stop this process, if there is just one machine $i < m$ left with a fractional block, and put this fractional block on machine $m$. Note that every machine $i$ received (fractional) jobs of (full) size at most $\rho w_i$, and the workload increased also by at most $\rho w_i$. The resulting job partition is called $P'_1, P'_2, \ldots, P'_m$ $P'_i = (L_i, S'_i)$.

For each $P'_i$ we define a configuration $\alpha_i$ in a recursive manner. Let $\lambda_i = \log_{(1+\delta)} \rho w_i$, and $\Lambda_i = \log_{(1+\delta)} w_i$. If $P'_i = \emptyset$, then let $\alpha_i$ be the empty configuration (see (C5)). Otherwise we calculate the unique $\mu_i$ s.t. (C5) holds for $L_i$. (1) implies $\rho w_i < (1+\delta)^{\mu_i}$, that is, $\lambda_i < \mu_i$. On the other hand, since $L_i$ contains at least one job $p$ s.t. $\frac{\delta \cdot |L_i|}{(1+\delta)^2} < \overline{p} \le w_i = (1+\delta)^{\Lambda_i}$, we have $\mu \le \Lambda_i$. For $m$ we define $\mu_m := \mu_{m-1}$.

The jobs in $P'_i$ are either blocks of size $\rho w_i$, or have size in $(\rho w_i, w_i]$. Let $\vec{n}^o$ of $\alpha_i$ be the null vector (cf. (V1)) if $i = 1$, and be the second size vector in $\alpha_{i-1}$ scaled to $w_i$ if $i > 1$. The $\vec{n}^1$ of $\alpha_i$ we can construct so that $L_{\alpha_i} = L_i$, and $S_{\alpha_i} = S'_i$. Here we exploit that every job in $S'_i$ is at most $\delta |L_i|$, and so it is in the $l$th class where $l \le \mu + 1$; similarly, that every job in $L_i$ is in some class $l \ge \mu$; and finally, that (A2) facilitates the consistent definition of the size vector coordinates. In the end, we can define a double configuration $(\alpha_{m-2}, \alpha_{m-1})$ consistent with (V4), since the partition fulfils (A) or (B).

Now as long some $\alpha_i$ exists, for which $n^1_\lambda > \max\{n^o_\lambda, \left\lfloor \frac{\sum_{l \le \lambda} |C_l|}{\rho w_i} \right\rfloor - 1\}$, (see (V3)), we 'put' tiny blocks from $P'_i$ to the set $P'_m$ (by correcting the $\alpha_i$). It is easy to see that for every valid magnitude we put at most two blocks to $m$, and the sum of these is at most $2\rho w_m$.

Clearly, the vertices $v_i = (II, i, \alpha_i)$ exist in $V_{II}$, and form an $m$-path in level II, as easily follows from the graph definition. We increased every workload $i \ne m$ by $\mathcal{O}(\delta) \cdot OPT$, so the theorem follows. □

## 5 The deterministic algorithm

This section describes the deterministic monotone algorithm, in form of two procedures (Sections 5.1 and 5.2), and the main algorithm PTAS (Section 5.3). In Section 5.4 we prove the monotonicity of the PTAS. We will make use of an arbitrary fixed total order $\prec$ over the set of all configurations $\alpha$, such that configurations of smaller total workload $|\alpha|$ are smaller according to $\prec$.

### 5.1 Computing an optimal path in $\mathcal{H}$

Procedure OPTPATH (see Figure 1) is a common dynamic programming algorithm that finds an $m$-path of minimum makespan in $\mathcal{H}_I$. However, we do not simply proceed from left to right over the $m$ graph layers, but select an optimal path from the first layer to every node in $V_I$, and similarly, an optimal path from layer $m$ to each node in $V_{II}$. Finally, we test each vertex in $V_{II}$ to provide a potential switch vertex (i.e., we find optimal paths leading to the switch vertex from both end-layers). When the makespan of two prefix (or suffix) paths is the same, we break ties according to $\prec$. We choose a switch vertex $v_k$ providing optimum makespan, and of maximum possible $k$. The case $k = m - 2$ needs careful optimization. Roughly, we choose *deterministically* by some fixed order of the configuration triples $(\alpha_{m-2}, \alpha_{m-1}, \alpha_m)$, but minimize the makespan on the last three machines by redistributing the tiny blocks. The flexibility provided by three machines with tiny blocks, facilitates monotone allocation in this degenerate case as well.

A pseudo-code of OPTPATH is presented in Figure 1. Observe, that by the definition of $M()$ and $opt()$ values, $M(v_k) = M(\mathcal{Q})$. Moreover, since in each case the pointers $pred()$ and $succ()$ determine an incoming, and an outgoing path of minimum makespan, respectively, we can make the following observation:

*Observation.* For all $v \in V_{II}$, the value $M(v)$ is the minimum makespan over all $m$-paths having vertex $v$ as switch vertex. Consequently, $M(v_k) = M(\mathcal{Q})$ is the minimum makespan over all $m$-paths

**Procedure 1** OPTPATH

Input: The directed graph $\mathcal{H}_I$.
Output: The optimal $m$-path $\mathcal{Q} = (v_1, \ldots, v_m)$ of $\mathcal{H}_I$.

1. for every double vertex $v'_{m-2} = (\alpha_{m-2}, \alpha_{m-1}) \in V_{II}$ do

    $opt(v'_{m-2}) := \max\{f(v_{m-2}), f(v_{m-1}), f(v_m)\}$ (where $\alpha_{m-1}$ determines $\alpha_m$, see Section 4)

    $M(v'_{m-2}) := \max\{\frac{|\alpha_{m-2}|}{s_{m-2}}, \frac{|\alpha_{m-1}|}{s_{m-1}}, \frac{|\alpha_m|}{s_m}\}$;

    for $i = m - 3$ downto 1 do

    for every $v = (d, i, \alpha) \in V_{II}$ do

    (i) $succ(v) := w$, if $opt(w) = \min\{opt(y) \mid (v, y) \in E\}$, and among such vertices of minimum $opt()$ the configuration $\alpha$ of vertex $w$ is minimal wrt. $\prec$.

    (ii) $opt(v) := \max\{f(v), opt(succ(v))\}$;
    $M(v) := \max\{\frac{|\alpha|}{s_i}, opt(succ(v))\}$.

2. for every $v = (d, 1, \alpha) \in V_I$ do $\quad opt(v) := f(v)$;

    for $i = 2$ to $m - 2$ do

    for every $v = (d, i, \alpha) \in V_I \cup V_{II}$ do

    (i) $pred(v) := w \in V_I$, if $opt(w) = \min\{opt(y) \mid y \in V_I, (y, v) \in E\}$, and among such vertices of minimum $opt()$ the configuration $\alpha$ of vertex $w$ is minimal wrt. $\prec$.

    (ii) if $v \in V_I$ then $opt(v) := \max\{f(v), opt(pred(v))\}$;
    if $v \in V_{II}$ then $M(v) := \max\{M(v), opt(pred(v))\}$.

3. select an optimal *switch vertex* $v_k = (II, k, \alpha_k) \in V_{II}$, by the following objectives:

    (i) $M(v_k) = \min\{M(v) \mid v \in V_{II}\}$;

    (ii) the layer $k$ is maximum over all $v$ of minimum $M(v)$;

    (iii) if $k = m - 2$, then among all double vertices $v'_{m-2} = (\alpha_{m-2}, \alpha_{m-1})$ of minimum $M(v'_{m-2})$ (and hidden configuration $\alpha_m$), select $v_k = v'_{m-2}$ by the following objectives:

    (a) keep the order (A) (B) (cf. Section 4);

    (b) in case of (A), select an $(\tilde{\alpha}_{m-2}, \tilde{\alpha}_{m-1}, \tilde{\alpha}_m, (T_{\alpha_{m-2}} \cup T_{\alpha_{m-1}} \cup T_{\alpha_m}))$ (i.e., with a common pool of tiny blocks) by some predefined ordering, then minimize the highest finish time of $\frac{|\alpha_{m-2}|}{s_{m-2}}, \frac{|\alpha_{m-1}|}{s_{m-1}}$, and $\frac{|\alpha_m|}{s_m}$, then minimize the *second* highest finish time among them (by redistributing the tiny jobs);

    (c) in case of (B), select an $(\alpha_{m-2}, \alpha_{m-1}, \alpha_m)$ by some predefined ordering, but so that $|\alpha_{m-1}| + |\alpha_m|$ is maximized;

    (iv) if $k \leq m - 3$ then the configuration $\alpha_k$ is minimal wrt. $\prec$ over all $v$ of minimum $M(v)$ in layer $k$;

4. for $i = k - 1$ downto 1 do $\quad v_i := pred(v_{i+1})$;

    for $i = k + 1$ to $m$ do $\quad v_i := succ(v_{i-1})$;

    $\mathcal{Q} := (v_1, \ldots, v_k, \ldots, v_m)$.

Figure 1: Procedure OPTPATH finds an $m$-path $\mathcal{Q}$ of minimum $M(\mathcal{Q})$ in the directed graph.

of $\mathcal{H}$.

## 5.2 Constructing the final job allocation

Once an optimal $m$-path is found, we have to allocate the jobs of $P_I$ to the machines. This is obvious for jobs that appear individually in some configuration of the path, but we need an accurate description of how the *tiny* jobs are distributed, given the block representation. Procedure PARTITION is detailed in Figure 2. Importantly, depending on whether the switch machine $k$ is filled high (above $(1 - \epsilon/2)$ times the makespan) or low, it gets filled with tiny jobs below, resp. above $|\alpha_k|$. This, again, will play an important role when showing monotonicity. Distributing the tiny jobs when $k = m - 2$, is a slightly more subtle procedure, operating with the same principle (a *low* machine is filled over $|\alpha_i|$). In general, having a careful look at PARTITION, one can see that the machines never get filled above the makespan of the input path $\frac{|Q_i|}{s_i} \leq M(\mathcal{Q})$. This is trivial for machines without tiny jobs, and follows from the definition of finish time with the extra tiny block, for other machines.

Procedure PARTITION is shown in Figure 2. The next two lemmas characterize properties of the output job-partition, that are essential for proving Theorem 4.

**Lemma 4.** *If $\mathcal{Q} = (v_1, \ldots, v_m)$ is the input path to procedure* PARTITION*, then the output $Q_1, \ldots, Q_m$ is, indeed, a partition of $P_I$, and the induced allocation $(Q_1, \ldots, Q_m)$ is canonical with the choice $L_i := L_{\alpha_i}$.*

*Proof.* The sets $\tilde{\alpha}_i$ are non-intersecting, as follows from the Definitions 8 and 9, and (E1). The set $T$ distributed for last, contains exactly the (tiny) jobs missing from $\bigcup_{i=1}^{m} \tilde{\alpha}_i$, so we really have a partition of $P_I$.

We claim that $(L_{\alpha_i}, S_{\alpha_i})$ is a $\delta$-division of $Q_i$: The sets $(L_{\alpha_i}, S_{\alpha_i} \setminus T_{\alpha_i})$ form a $\delta$-division of each $\tilde{\alpha}_i$, according to Lemma 3. We show that the tiny jobs allocated to any machine $i$ have size at most $\rho w_i$. Note that $W$ denotes the total size of tiny blocks in $\bigcup_{h=1}^{i} \alpha_h$ in the $i$th round of step 3. Let $\tau$ denote the total size of jobs tiny for $w_i$ that were allocated (as non-tiny) in $\bigcup_{h=1}^{i-1} \tilde{\alpha}_h$. (S5) implies that if $n_\lambda^1$ is in the second size vector of $\alpha_i$, then $W + \tau < n_\lambda^1 \cdot \rho w_i + \rho w_i$. Moreover, (V3) implies that $n_\lambda^1 \cdot \rho w_i \leq \sum_{l \leq \lambda} |C_l| - \rho w_i$, whenever $n_\lambda^o < n_\lambda^1$. So, $W + \tau < \sum_{l \leq \lambda} |C_l|$, so there are enough tiny jobs in $T$ from the classes $l \leq \lambda$ to fulfil (ii) of step 3b.

Furthermore, (A1) holds by (E2), and (A2) holds because we defined the configurations and $Scale()$ consistent with (A2), and tiny jobs are allocated in increasing order. □

**Lemma 5.** *If $\mathcal{Q} = (v_1, \ldots, v_m)$ is the input path, then for the output $Q_1, \ldots, Q_m$ of* PARTITION $\frac{|\alpha_i| - 6\rho w_i}{s_i} \leq \frac{|Q_i|}{s_i} \leq M(\mathcal{Q})$ *for every machine $i$.*

*Proof.* For $i < k$, $\alpha_i$ contains no tiny jobs, and $Q_i = \alpha_i$. So, in this case $\frac{|Q_i|}{s_i} = \frac{|\alpha_i|}{s_i} = f(v_i) \leq M(\mathcal{Q})$.

In PARTITION, the variable $W_i$ stands for the work of tiny blocks assigned to $i$ by $\alpha_i$. As for $i = k$, $Q_k$ is the first set (possibly) containing tiny jobs, so it certainly receives an amount of $W_k \pm \rho w_k$ tiny jobs, and not more than $W_k$ in case of HIGH-$k$.

It is straightforward to check that if $k \leq m - 3$, then the last machine with tiny blocks ($i \in \{m-1, m\}$) receives at most $W_i$, and at least $W_i - 6\rho w_i$ work (due to the estimate $n_\lambda^1 = \left\lceil \frac{\sum_{l \leq \lambda} |C_l|}{\rho w} \right\rceil + 3$ in the hidden configuration $\alpha_m$). The other machines $i > k$ get tiny jobs of total size $W_i \pm \rho w_i$. Note that the $+\rho w_i$ overhead was calculated in these machines' finish times $f(v)$, and indirectly in the makespan $M(\mathcal{Q})$.

Finally, in case $k = m - 2$, only a low machine $i \in \{m - 2, m - 1, m\}$ may receive more work than $|\alpha_i|$; and each machine receives at least $W_i - 6\rho w_i$ work of tiny jobs. □

**Procedure 2** PARTITION

Input: The job set $P_I$, and an $m$-path $\mathcal{Q} = (v_1, \ldots, v_m)$ with switch vertex $v_k$ in the graph $\mathcal{H}_I$.
Output: A partition $Q_1, Q_2, \ldots, Q_m$ of the set $P_I$.

Case LOW-$k$ : $|\alpha_k|/s_k \leq (1 - \epsilon/2) \cdot M(\mathcal{Q})$
Case HIGH-$k$ : $|\alpha_k|/s_k > (1 - \epsilon/2) \cdot M(\mathcal{Q})$.

1. for $i = 1$ to $m$ do

    $Q_i := \tilde{\alpha}_i$;

2. let $T = \{t_1, t_2, t_3, \ldots, t_\mu\} = P_I \setminus \bigcup_{i=1}^{m} Q_i$, so that the jobs $t_j$ are in increasing order, and this order corresponds to the order of jobs in each class.

3a. if $k = m - 2$, then start with an allocation of tiny jobs to $\{m-2, m-1, m\}$ (in the given order) s. t. each machine $i$ gets at most $|T_{\alpha_i}|$ amount of tiny jobs (this is doable, because the total number of tiny blocks is overestimated by 3 blocks in $\alpha_m$)

    let $M = \max\{\frac{|\alpha_{m-2}|}{s_{m-2}}, \frac{|\alpha_{m-1}|}{s_{m-1}}, \frac{|\alpha_m|}{s_m}\}$;

    call $i \in \{m-2, m-1, m\}$ *low*, if $|\alpha_i|/s_i \leq (1 - 2\epsilon/3) \cdot M$, and *high* if $|\alpha_i|/s_i \geq (1 - \epsilon/2) \cdot M$;

    Correct the partition of tiny jobs (with keeping the job order) so that

    (i) if there is one low machine $i$, and two high machines, then $i$ receives at least $|\alpha_i|$ work;

    (ii) if there are two non-high machines, then both receive at least $6\rho w$ work of tiny jobs.

3b. if $k \leq m - 3$, then let $W = 0$ and $r = 0$;

    for $i = k$ to $m - 1$ do

        given $\alpha_i = (w, \mu, \vec{n}^o, \vec{n}^1)$ and $\lambda = \log \rho w$, let $W_i := (n_\lambda^1 - n_\lambda^o) \cdot \rho w$;

    (i) $W := W + W_i$;

    (ii) if HIGH-$k$ then let $u$ be the maximum index in $T$ so that $\sum_{j=1}^{u} t_j \leq W$;
        if LOW-$k$ then let $u$ be the minimum index in $T$ so that $\sum_{j=1}^{u} t_j \geq W$;

    (iii) $Q_i := Q_i \cup \{t_{r+1}, t_{r+2}, \ldots, t_u\}$;

    (iv) $r := u$.

    $Q_m := Q_m \cup \{t_{r+1}, t_{r+2}, \ldots, t_\mu\}$.

Figure 2: Procedure PARTITION allocates the jobs based on path $\mathcal{Q}$ output by OPTPATH.

The next lemma is a descriptive characterization of the final output allocation $P_1, \ldots, P_m$ to be readily used in the monotonicity proof.

**Lemma 6.** *Let $P_I$ denote the set of input jobs and $s$ be the vector of rounded speeds. Let $\mathcal{Q} = (v_1, \ldots, v_m)$ be the path output by* OptPath *with configurations $(\alpha_1, \ldots, \alpha_m)$ in the vertices, and $v_k = (II, k, \alpha_k)$ be the switch vertex of the path. If $P_1, \ldots, P_m$ is the final partition output by* Ptas, *then*

(a) *for $i < k$, $P_i = Q_i = \alpha_i$.*

*Moreover, if $k \leq m - 3$, then*

(b) *for $i > k$, $\frac{|P_i|}{s_i} \geq (1 - 6\delta) \cdot M(\mathcal{Q})$;*

(c) *for $k$ either $P_k = Q_k$ or (b) holds.*

*Proof.* Point (a) is obvious, since $S_{\alpha_i} = \emptyset$ for every vertex $v_i \in V_I$, and by step 1. of Partition, $L_{\alpha_i} = \alpha_i = Q_i$. Moreover, by (E2) $L_{\alpha_i}$ has the $i$th smallest weight, so $Q_i = P_i$.

Let $M = M(\mathcal{Q})$. For $i > k$ we first claim that $|\alpha_i| \geq (1 - 2\delta) \cdot M \cdot s_i$. Assuming the contrary, $|\alpha_i| < (1 - 2\delta) \cdot M \cdot s_i$, we could put a job $p \in S_{\alpha_k}$ to machine $i$. Since by Proposition 3, $\overline{p} \leq \delta |L_{\alpha_k}| < \delta |\alpha_i|$ holds, this would increase $|\alpha_i|$ to at most $|\alpha_i'| \leq (1 + \delta)|\alpha_i| < (1 - \delta) \cdot M \cdot s_i$. The same would hold if $i$ increased its number of blocks by one. Notice that the workload of other machines also changes if they have jobs from the class of $p$ (because we defined $\mathcal{H}$ consistent with (A2)), however now they just get smaller jobs of the same class. Applying inequality (1), we obtain $f(v_i') \leq \frac{|\alpha_i'| + \rho w_i}{s_i} \leq \frac{|\alpha_i'|(1 + \delta)}{s_i} \leq M$.

We decreased $|\alpha_k|$ (or even found a valid path with switch vertex $k+1$ if no small jobs remained on $k$), without having increased the makespan, so $\mathcal{Q}$ was not optimal. Now Lemma 5 and inequality (1) yields $|Q_i| \geq |\alpha_i| - 6\rho w \geq |\alpha_i|(1 - 4\delta) > (1 - 6\delta) \cdot M \cdot s_i$. Since this holds for every $i > k$, it also holds for the ordered partition, which proves (b).

In order to see (c), notice that (a) implies $P_k \neq Q_i$ for any $i < k$. For $i > k$ we showed above that $|Q_i| \geq (1 - 6\delta) \cdot M \cdot s_i$. If the same holds for $i = k$, then (b) holds for $i = k$ as well; otherwise $|Q_k| = \min_{i \geq k} |Q_i|$, so $P_k = Q_k$. □

**Corollary 1.** *In step 5. of* Ptas, *the sets $Q_i$ are permuted only among machines $i \geq k$ of equal rounded speed. Consequently, $|P_i|/s_i \leq M(\mathcal{Q})$ for all $i$.*

*Proof.* Assume that $k \leq m - 3$, and $k \leq h < i$. If $s_i > s_h$ then by Lemma 6 we have that $|Q_i| \geq (1 - 6\delta) \cdot M(\mathcal{Q}) \cdot s_i \geq (1 - 6\delta) \cdot M(\mathcal{Q}) \cdot (1 + \epsilon) \cdot s_h > M(\mathcal{Q}) \cdot s_h \geq |Q_h|$. Now let $k = m - 2$, $h, i \in \{m - 2, m - 1, m\}$, and $s_h < s_i$. Since $\tilde{\alpha}_i$ is increasing, $Q_h > Q_i$ could happen only if $h$ receives tiny jobs, and has finish time about a factor of $(1 + \epsilon)$ higher than $i$. However, then either the highest finish time, or the second highest finish time among the machines $\{m - 2, m - 1, m\}$ would not be minimized by redistributing the tiny jobs, as required by OptPath. □

Corollary 1, and Theorem 2 imply following the upper bound:

**Theorem 3** *For arbitrary input $I$, and any given $0 < \epsilon \leq 1$, the deterministic algorithm* Ptas *outputs a $(1 + 3\epsilon)$-approximate optimal allocation in time $Poly(n, m)$.*

*Proof.* By Theorem 2, for $\delta \ll \epsilon$ the optimal path $\mathcal{Q}$ in $\mathcal{H}_I$ has makespan $M(\mathcal{Q}) < (1 + \epsilon)OPT$, and by Corollary 1 this remains an upper bound for the makespan of the output. Since Ptas rounds the input speeds to integral powers of $(1 + \epsilon)$, we obtain an overall approximation factor of at most $(1 + \epsilon)^2 \leq (1 + 3\epsilon)$.

**Algorithm 3** PTAS

Input: machine speeds $\sigma_1 \leq \sigma_2 \leq \ldots \leq \sigma_m$, and job set $P_I = \{p_1, p_2, \ldots, p_n\}$, desired precision $\epsilon$.
Output: A partition $P_1, P_2, \ldots, P_m$ of $P_I$.

1. for each $i \in [1, m]$, round the speed $\sigma_i$ up to the nearest power of $(1 + \epsilon)$;

2. based on the jobs $P_I$, rounded speeds $s_1 \leq s_2 \leq \ldots \leq s_m$, and an appropriate $\delta \ll \epsilon$, construct the graph $\mathcal{H}_I$;

3. run Procedure OPTPATH in order to obtain the optimal $m$-path $\mathcal{Q} = (v_1, \ldots, v_m)$;

4. from $\mathcal{Q}$ compute the partition $Q_1, Q_2, \ldots, Q_m$ by Procedure PARTITION;

5. let $P_1, P_2, \ldots, P_m$ be the sets of $\{Q_1, Q_2, \ldots, Q_m\}$, sorted by increasing order of weight $|Q_i|$; output $P_1, P_2, \ldots, P_m$.

Figure 3: The deterministic monotone PTAS.

In order to prove the running time bound, we show that for constant $\epsilon$, step 2. of PTAS can be computed efficiently. The number of graph vertices $|V|$ is $\mathcal{O}(m \cdot A)$, where $A$ denotes the number of different configurations. Every configuration (including the double configurations of triple length) is determined by $\mathcal{O}(\log_{(1+\delta)} 1/\rho) = \mathcal{O}((1/\delta) \cdot \log 1/\delta)$ coordinates, each of which (including $\mu$) may take at most $n + 1$ different values, so $|A| = n^{\mathcal{O}((1/\delta) \cdot \log 1/\delta)}$. Finally, for any $v, v' \in V$ deciding whether $(v, v') \in E$, takes time linear in the number of these coordinates. Thus for fixed $\delta$, step 2. is $poly(n, m)$, and steps 3., 4. and 5. are obviously polynomial, which completes the proof. □

### 5.3 The monotone PTAS

The monotone PTAS is presented in Figure 3. A substantial property of the output is that workloads $Q_i$ without small jobs do not get permuted in step 5. of PTAS. This is due to the fact that the sets $L_i$ of large jobs are increasing by (E2) (resp. that $\tilde{\alpha}_i$ is increasing in case of (A)). On the other hand, machines having small jobs, except for the switch machine, have finish time close to the makespan $M(\mathcal{Q})$ (resp. finish time of small difference in (A)). As a consequence, we obtain that in step 5. the sets $Q_i$ are permuted only among machines $i \geq k$ of *equal* rounded speed $s_i$. Therefore, even for the permuted workloads, $|P_i|/s_i \leq M(\mathcal{Q})$ holds for all $i$. This, in turn, together with Theorem 2 implies the following:

**Theorem 3.** *For arbitrary input $I$, and any given $0 < \epsilon \leq 1$, the deterministic algorithm PTAS outputs a $(1 + 3\epsilon)$-approximate optimal allocation in time $Poly(n, m)$.*

### 5.4 Monotonicity

Our main result is the following theorem.

**Theorem 4.** *Algorithm PTAS is monotone.*

*Proof.* Assume that machine $i$ alone decreased its speed $\sigma_i$ to $\sigma_i'$ in the input. If the vector of rounded speeds $(s_1, s_2, \ldots, s_m)$ remains the same, then the deterministic PTAS outputs the same allocation, and $i$ receives the same, or smaller workload, since the output workloads $P_1, P_2, \ldots, P_m$ are in increasing order. Assuming that the rounded speed $s_i$ decreased as well, it is enough to consider the special case when $i$ is the first (smallest index) machine of rounded speed $s_i = (1+\epsilon)$ in input $I(P, s)$, and after reducing its speed, it becomes the last (highest index) machine of rounded speed 1 in input $I'(P, s')$. Since the workloads in the final allocation $P_1, \ldots, P_m$ are ordered, this implies monotonicity for every 'one-step' speed change (like $(1+\epsilon) \to 1$). Monotonicity in general can then be obtained by applying such a step repeatedly. Note that for both inputs the algorithm constructs the same graph, independently of the speed vector. We assume that with inputs $I$, and $I'$ OPTPATH outputs $\mathcal{Q} = (v_1, \ldots, v_m)$, and $\mathcal{Q}' = (v_1', \ldots, v_m')$, where the switch vertices have index $k$ and $k'$, respectively. Finish time, makespan, etc. wrt. the *new* speed vector $s'$ are denoted by $f'(), M'()$ etc. We prove that $|P_i| \geq |P_i'|$.

We start with a simple observation. Since we increased a machine speed, it follows from the definition of makespan that for any path $\mathcal{R} = (v_1, v_2, \ldots, v_r)$ in layer $V_I$, or in layer $V_{II}$, and for any $m$-path, $M'(\mathcal{R}) \geq M(\mathcal{R})$. Similarly, for any vertex $v$, $opt'(v) \geq opt(v)$, and for any $v \in V_{II}$ $M'(v) \geq M(v)$ (cf. Procedure OPTPATH). Obviously, also the optimum makespan over all $m$-paths could not decrease. We elaborate on the subtle case of $k = m-2$ in a separate lemma; in what follows, we assume $k \leq m-3$.

CASE 1: $M'(\mathcal{Q}) > M(\mathcal{Q})$

In this case machine $i$ with the new rounded speed $s_i' = 1$, becomes a bottleneck in path $\mathcal{Q}$. That is, $M'(\mathcal{Q}) = f'(v_i) = \frac{|\alpha_i|(+\rho w)}{1}$.

If $i < k$, then $\alpha_i = P_i$, so the machine received exactly $|\alpha_i|$ work with speed $s_i$, and now $\mathcal{Q}$ is a path with makespan $|\alpha_i|$, so by Corollary 1, and by the optimality of $\mathcal{Q}'$ we have $|P_i'| = |P_i'|/s_i' \leq M'(\mathcal{Q}') \leq M'(\mathcal{Q}) = |\alpha_i| = |P_i|$.

Recall that $s_i = (1+\epsilon)$. Let us introduce the notation $B \stackrel{\text{def}}{=} (1-6\delta) \cdot M(\mathcal{Q})$ for the lower bound of Lemma 6. Assume now that $i \geq k$ and $|P_i|/(1+\epsilon) \geq B$.

Due to (E2), for the job partition $Q_1, \ldots, Q_m$ (before ordering the sets by size), it holds that $|Q_h| \geq |L_{\alpha_i}|$ for every $h \geq i$. Therefore, for the $i$th largest set $P_i$, we have $|P_i| \geq |L_{\alpha_i}|$, and so $|P_i| \geq \max\{|L_{\alpha_i}|, (1+\epsilon) \cdot B\}$.

We modify the path $\mathcal{Q}$ and construct a new path $\mathcal{Q}''$ by 'putting' small jobs from $S_{\alpha_i}$ (of machine $i$) onto machine $i+1$, until the moved jobs have total weight of at least $7\delta \cdot (1+\epsilon) \cdot M(\mathcal{Q})$, or $S_{\alpha_i}$ becomes empty. For the new finish time (using (1) we have $f'(v_i'') \leq \max\{|L_{\alpha_i}|, (1+\epsilon) \cdot M(\mathcal{Q})(1-7\delta) + \rho w\} \leq \max\{|L_{\alpha_i}|, (1+\epsilon) \cdot B\} \leq |P_i|$. If $i = m$, then we put only *tiny* blocks of the common magnitude $w_{m-1}$ onto $m-1$, and use $|\tilde{\alpha}_m|$ instead of $|L_{\alpha_i}|$ in the calculation.

We claim that with speed $s_i' = 1$ machine $i$ is a bottleneck machine in both paths $\mathcal{Q}$ and $\mathcal{Q}''$. For $i > k$, it follows from the optimality of $\mathcal{Q}$ (as shown in the proof of Lemma 6) that $|\alpha_i| \geq (1+\epsilon) \cdot M(\mathcal{Q})(1-2\delta)$. (Note that, as a consequence, $M'(\mathcal{Q}) = M(\mathcal{Q})$ is possible only if $i \leq k$.) For $i = k$ it follows from PARTITION and by assumption on $|P_i|$ that $|\alpha_i| \geq |Q_i| \geq (1+\epsilon) \cdot M(\mathcal{Q})(1-6\delta)$. After removing the small jobs, in the new path $f'(v_i'') = M'(\mathcal{Q}'')$, because $f'(v_i'') \geq (1+\epsilon) \cdot M(\mathcal{Q})(1-13\delta) \geq M(\mathcal{Q})$. Thus, $f'(v'')$ is an upper bound on the new optimal path-makespan, and it is less than $|P_i|$.

By Lemma 6, it remains to consider the case $i = k$, and $P_i = Q_i$. Given $M'(\mathcal{Q}) > M(\mathcal{Q})$, we have $M'(\mathcal{Q}') \leq M'(\mathcal{Q}) = |\alpha_i|$ as an upper bound on $|P_i'|$. Assuming that for $i = k$ Low-$k$ holds with speed $s_i$, $|P_i| = |Q_i| \geq |\alpha_i|$ by PARTITION 3b, and we are done. Assuming HIGH-$k$, $|P_i| = |Q_i| \geq \max\{|\tilde{\alpha}_i|, |\alpha_i| - \rho w_i\}$. On the other hand, $M'(\mathcal{Q}) = |\alpha_i| > (1-\epsilon/2)M(\mathcal{Q}) \cdot (1+\epsilon) > (1+\epsilon/3)M(\mathcal{Q})$. By putting one tiny block onto the next machine (if there are any), we still obtain a path $\mathcal{Q}''$, with makespan $M'(\mathcal{Q}'') = |\alpha_i''| = \max\{|\tilde{\alpha}_i|, |\alpha_i| - \rho w_i\} > M(\mathcal{Q})$, and we are done.

CASE 2. $M'(\mathcal{Q}) = M(\mathcal{Q})$, and $i \leq k$.

If $M'(\mathcal{Q}) = M(\mathcal{Q})$, and $\mathcal{Q} = \mathcal{Q}'$, then the output of PARTITION can only be different if $i = k$. Since the makespan did not change, and $s_k$ decreased, the change is from LOW-$k$ to HIGH-$k$, and machine $i = k$ receives less work with $s'_k$, by PARTITION 3b. If, on the other hand, PARTITION outputs the same partition, then every machine gets the same workload.

Suppose $M'(\mathcal{Q}) = M(\mathcal{Q})$, but $\mathcal{Q} \neq \mathcal{Q}'$. Since $\mathcal{Q}$ has minimum makespan, $M'(\mathcal{Q}') = M'(\mathcal{Q})$. We claim that also $k' = k$. Otherwise $\mathcal{Q}'$ would have been better than $\mathcal{Q}$ for input $s$ as well, because $M(\mathcal{Q}') \leq M'(\mathcal{Q}') = M(\mathcal{Q})$ and $k' > k$. Similarly, also $v_k = v'_k$, otherwise $\alpha' \prec \alpha$ would hold, and $\mathcal{Q}'$ would have been better for input $s$ as well.

Now, since $\mathcal{Q} \neq \mathcal{Q}'$, a maximum $h < k$ exists so that $v_h \neq v'_h$. This means that $pred(v_{h+1}) \neq pred'(v_{h+1})$. If $v'_h$ was preferred in $\mathcal{Q}'$ because $\alpha'_h \prec \alpha_h$, then $opt(v_h) < opt(v'_h) \leq opt'(v'_h) \leq opt'(v_h)$. The first inequality holds, otherwise $v'_h = pred(v_{h+1})$ would have been the choice. The second holds for every vertex. The third holds, otherwise $v_h = pred'(v_{h+1})$ would have been the choice of OPTPATH. Similarly, if $v'_h$ was preferred in $\mathcal{Q}'$ because $opt'(v'_h) < opt'(v_h)$, then $opt(v_h) \leq opt(v'_h) \leq opt'(v'_h) < opt'(v_h)$. In both cases we obtained $opt(v_h) < opt'(v_h)$. Recall that $opt(v_h)$ is the optimum makespan over all paths leading to $v_h$ from layer 1. This could strictly increase only if $i \leq h$, and $i$ with workload $\alpha_i = P_i$ and speed $s'_i = 1$ became a bottleneck machine in $(v_1, v_2, \ldots, v_h)$. Therefore, $|P'_i| \leq opt'(v'_h) \leq opt'(v_h) \leq \frac{|\alpha_i|}{s'_i} = |P_i|$.

**Lemma 7.** *If on input $I(P,s)$, for the output path $\mathcal{Q}$ of OPTPATH the switch index is $k = m - 2$, then $|P_i| \geq |P'_i|$.*

*Proof.* For $i < k$ the proof is exactly the same as in CASES 1. and 2. of the theorem, since the structure of the output solution on machines $i \geq k$ does not affect that argument. In the rest of the proof we assume that $i \geq k = m - 2$. In this case, for $i < k$ clearly $Q_i = L_i$ is non-decreasing. Furthermore we make use of the following:

**Claim 1.** Step 3a. of PARTITION can be realized so that $|Q_{m-2}| \leq |Q_{m-1}| \leq |Q_m|$ holds.

The claim implies $Q_i = P_i$ for all $i \in [1, m]$, so no permutation in step 5. of PTAS takes place, which simplifies the monotonicity argment below.

To see the claim, observe that for every path where $k = m - 2$, $|\tilde{\alpha}_i|$ and $|\alpha_i|$ are non-decreasing (cf (A) and (B) in Section 4). It is not hard to see that, PARTITION can allocate the tiny jobs in increasing order to the machines so that $|Q_i|$ is also increasing, and Lemma 5 still holds. (Here we exploit that the number of machines with tiny jobs is constant.) Now consider 3a. of PARTITION. If machine $Q_i$ is increased (corrected) because of (i), then $i$ still gets much less work than any higher index machines; if $Q_i$ is increased because of (ii), then $i$ has about 6 tiny blocks whereas the other non-high machine gets at least 12 tiny blocks (and the bottleneck machine gets no blocks). Thus, no higher index machine can get less work than $i$, and the proof of the claim is completed.

We do not give a detailed proof for the cases (A)(i) and (B)(i). In both cases all tiny jobs are on the last two machines. These jobs can be allocated to machines $m-1, m$ in increasing order, so that the makespan on $(m-1, m)$ is optimized, and this optimized makespan can be computed *exactly* already during the path optimization. It can obviously be assumed in both (A) and (B) that $|\alpha_{m-2}| \leq |\alpha_{m-1}| \leq |\alpha_m|$. The proof is therefore analogous to the proof of case $i < k$.

Now suppose that $i \geq k = m - 2$, and for path $\mathcal{Q}$ (A)(ii) holds. Let $M = \max\{\frac{|\alpha_{m-2}|}{s_{m-2}}, \frac{|\alpha_{m-1}|}{s_{m-1}}, \frac{|\alpha_m|}{s_m}\}$. Assume first, that machine $i$ with speed $s_i$ was a non-low machine in 3a. of PARTITION. Then, with speed $s'_i = 1$ machine $i$ becomes a bottleneck in the subpath $(v_{m-2}, v_{m-1}, v_m)$, and maybe even in the path $\mathcal{Q}$. Moreover, by $\delta \ll \epsilon$, machine $i$ is still a bottleneck in a modified (sub)path $\mathcal{Q}''$, where we put 7 tiny blocks, or *all* tiny blocks from $i$ to another one of the last three machines. That

is, the makespan of $\mathcal{Q}''$ (resp. the local makespan on $(m-2, m-1, m)$) is $\max\{|\tilde{\alpha}_i|, |\alpha_i| - 7\rho w_i\}$, which is an upper bound on $|P_i'|$, and a lower bound on $|P_i|$, by Lemma 5.

Now assume that $i$ was a low machine in 3a. of PARTITION.

If there was another non-high machine $i'$ then, by the optimization rules 3. (iii) of OPTPATH, only $i$ and $i'$ have tiny blocks in $\mathcal{Q}$. Having changed the speed to $s_i'$, we construct a path $\mathcal{Q}''$ by putting tiny blocks (when necessary) from $i$ to $i'$ so that their maximum finish time is minimized. If, with the optimized tiny blocks, the local makespan remains $M$ then $\mathcal{Q}' = \mathcal{Q}''$ is the new output solution. (Any path preferred to $\mathcal{Q}''$ would have been preferred with speed $s_i$ as well.) Obviously, in $\mathcal{Q}''$ $i$ gets no more tiny blocks than in $\mathcal{Q}$, and in PARTITION $i$ receives a subset of the tiny jobs that it received with speed $s_i$. If the local makespan increases then $i$ becomes a (local) bottleneck so it has at most 6 blocks in $\mathcal{Q}''$ (further blocks could be put on $i'$). Thus, $|\tilde{\alpha}_i| + 6\rho w_i$ is an upper bound on the new local makespan and also on $|P_i'|$; whereas, by PARTITION 3a (ii) it is a lower bound on $|P_i|$.

Suppose that $i$ was a low machine, and the other two were both high machines. It easily follows from 3 (iii) of OPTPATH that $i$ has nearly all tiny blocks with speed $s_i$; whereas by 3a (i) of PARTITION $|P_i| \geq |\alpha_i|$ holds. Consider now the same path $\mathcal{Q}$ with speed $s_i'$. If $i$ becomes a local bottleneck (i.e., among $\{m-2, m-1, m\}$) then $|\alpha_i|$ is an upper bound on the optimal (local) makespan of $\mathcal{Q}'$, so that $|P_i'| \leq |\alpha_i|$, and we are done. If $i$ has the second highest finish time among $\{m-2, m-1, m\}$, then the output path remains basically the same (by 3 (iii) of OPTPATH), possibly optimizing the second finish time by removing blocks from $i$. $i$ is not a low machine anymore, and $|P_i'| \leq |\alpha_i|$. If $i$ has still the lowest finish time, then the output path is the same, and in PARTITION $i$ gets the same set, or a subset of his previous jobs in case he becomes a non-low machine.

Finally, we turn to the case when $i \geq k = m-2$, and for path $\mathcal{Q}$ (B) (ii) holds. We claim that the optimality of $\mathcal{Q}$ implies that machines $m-1$ and $m$ have finish time at least $(1 - 6\delta)M(\mathcal{Q})$ (cf. Lemma 6). Based on this, one can easily verify that the allocation of PARTITION is essentially the same as in the case $k \leq m-3$, and the monotonicity proof is analogous.

In the rest of the proof, we show that the claim holds. Recall that in case (B) (ii), OPTPATH maximizes $|\alpha_{m-1}| + |\alpha_m|$. As long as machine $m-3$ has at least one small job, the same argument as in Lemma 6 can be used: if $m-1$ or $m$ are not filled enough, then we could shift a small job from $m-2$ to these machines, increasing $|\alpha_{m-1}| + |\alpha_m|$, without increasing $M(\mathcal{Q})$, a contradiction.

Assume that $m-2$ has no small jobs, that is, $\alpha_{m-2} = L_{m-2}$. We show that the *whole* job set $L_{m-2}$ has the size of at most that of a tiny job for $w_{m-1}$. By the condition (B), $\rho^2 w_{m-1} > w_{m-2}$. Since $w_{m-2}$ is the maximum job size in $L_{m-2}$, and by property (D2) of $\delta$-divisions, we have $w_{m-2} > \frac{\delta |L_{m-2}|}{(1+\delta)^2} > \rho |L_{m-2}|$. The latter inequality follows from $\rho < \delta/3$ in case $1 + \delta < \sqrt{3}$. Putting it together, we obtain $\rho^2 w_{m-1} > \rho|L_{m-2}|$, so $|L_{m-2}|$ is tiny for $w_{m-1}$. We define a path $\mathcal{Q}''$ as follows. We put all the jobs of $L_{m-2}$ to $m-1$, and shift the jobs of the corresponding classes to $m$, when necessary (i.e., when $m-1$ was filled, but $m$ not). Also, shift every workload $\alpha_i = L_i$ to the next machine for machines $i \leq m-3$. Finally, note that the new partition is canonical: the jobs in $L_{m-2}$ were the last large jobs in their class, now they become the first small jobs, so (A2) remains valid. This means that we found a path better than the optimum $\mathcal{Q}$, a contradiction. □

□